\documentclass[a4paper,11pt]{article}
\usepackage{jcappub}
\usepackage{bm}
\usepackage{float}
\usepackage{subfigure}
\usepackage{orcidlink}
\usepackage{CJK}

\makeatletter
\gdef\@fpheader{Published in JCAP}
\@addtoreset{equation}{section}

\makeatother

\title{Analytical sensitivity curves of the second-generation time-delay interferometry}

\author[a,1]{Chunyu Zhang~\orcidlink{0000-0002-4332-3729}\note{Corresponding author.}}

\affiliation[a]{Center for Gravitation and Cosmology, College of Physical Science and Technology, Yangzhou University, Yangzhou 225009, China}

\emailAdd{chunyuzhang@yzu.edu.cn}

\abstract{
Forthcoming space-based gravitational-wave (GW) detectors will employ second-generation time-delay interferometry (TDI) to suppress laser frequency noise and achieve the sensitivity required for GW detection.
We introduce an inverse light‑path operator $\mathcal{P}_{i_{1}i_{2}i_{3}\ldots i_{n-1}i_{n}}$, which enables simple representation of second‑generation TDI combinations and a concise description of light propagation.
Under the instantaneous equal-arm, equilateral triangular-constellation approximation, analytical expressions and high‑accuracy approximate formulas are derived for the sky‑ and polarization‑averaged response functions, noise power spectral densities (PSDs), and sensitivity curves of TDI Michelson, Sagnac-type ($\alpha,\beta,\gamma$), Monitor, Beacon, Relay, and fully symmetric Sagnac ($\zeta$) combinations, as well as their orthogonal $A, E, T$ channels.
The validity of the analytical results is confirmed via Monte Carlo integration with $10^{4}$ randomly sampled sky locations and polarization angles.
Our results show that:
(i) second‑generation TDIs have the same sensitivities as their first‑generation counterparts;
(ii) the $A, E, T$ sensitivities and the optimal sensitivity are independent of the TDI generation and specific combination;
(iii) the $A$ and $E$ channels have equal averaged responses, noise PSDs, and sensitivities,
while the $T$ channel has much weaker response and sensitivity at low frequencies;
(iv) except for the $(\alpha,\beta,\gamma)$ and $\zeta$ combinations and the $T$ channel, all sensitivity curves exhibit a flat section;
(v) the averaged response, noise PSD, and sensitivity of $\zeta$ scale with those of the $T$ channel.
These analytical and approximate formulations provide useful equal-arm benchmarks for instrument optimization and data-analysis preparatory studies for future space-based GW detectors.
}

\begin{document}

\maketitle	
\flushbottom

\section{Introduction}

The first direct detection of gravitational waves (GWs) from a binary black hole merger by the Advanced Laser Interferometer Gravitational-Wave Observatory (aLIGO) and Virgo marked the beginning of a new era in multi-messenger astronomy \cite{LIGOScientific:2016aoc,LIGOScientific:2016emj,LIGOScientific:2016vlm} and has provided unprecedented opportunities to explore the Universe and test fundamental physics \cite{Abbott:2016nmj,Abbott:2017vtc,Abbott:2017oio,TheLIGOScientific:2017qsa,LIGOScientific:2017ync,Abbott:2017gyy}. 
Since that groundbreaking discovery, the LIGO–Virgo–KAGRA (LVK) Collaboration has reported nearly 100 GW events \cite{LIGOScientific:2018mvr,LIGOScientific:2020ibl,KAGRA:2021vkt,LIGOScientific:2025slb}.
Ground-based detectors, such as Advanced LIGO \cite{Harry:2010zz,TheLIGOScientific:2014jea}, Advanced Virgo \cite{TheVirgo:2014hva} and Kamioka Gravitational Wave Detector (KAGRA) \cite{Somiya:2011np,Aso:2013eba}, operating in the $10-10^4$ Hz frequency band, normally detect stellar-mass binary mergers.
The proposed space-based detectors such as Laser Interferometer Space Antenna (LISA) \cite{Danzmann:1997hm,LISA:2017pwj}, TianQin \cite{Luo:2015ght}, and Taiji \cite{Hu:2017mde} are designed to cover the $10^{-4}-1$ Hz frequency band, while Deci-hertz Interferometer Gravitational Wave Observatory (DECIGO) \cite{Kawamura:2011zz} targets the $0.1-10$ Hz frequency band.
These space-based GW detectors will enable detections of GWs from massive black hole binary mergers, providing powerful opportunities to probe the fundamental properties of gravity and the astrophysical processes governing compact objects.

A space-based GW detector will consist of three spacecraft forming a nearly equilateral triangular array, exchanging coherent laser beams along arms millions of kilometers long.
The unequal and slowly time-varying arm lengths, with variations on the order of $\lesssim 10$ m/s due to the orbital motion around the Sun \cite{LISA:2017pwj}, 
make the suppression of laser frequency noise (more than 7 orders of magnitude larger than other secondary noises) extremely challenging.
Therefore, space-based detectors must implement a specific data-processing technique known as time-delay interferometry (TDI) to achieve
the requisite sensitivity.

TDI relies on constructing linear combinations of appropriately delayed one-way interspacecraft measurements.
The first-generation TDI considers three armlengths and assumes them to be constant \cite{Tinto:1999yr,Armstrong:1999,Estabrook:2000ef}, for which the three time-delay operators commute and form a commutative polynomial ring.
With a rigorous mathematical foundation called \textit{first module of syzygies}, 
any first-generation TDI combination can be written as a linear combination of four generators, $\alpha,\beta,\gamma$ and $\zeta$ \cite{Dhurandhar:2001tct,Tinto:2020fcc}.

The 1.5-generation TDI considers the Sagnac effect that the up and down optical paths are unequal, and assumes the six armlengths to be constant.
With a rigorous mathematical foundation, any 1.5-generation TDI combination can be written as a linear combination of six generators \cite{Tinto:2003vj,Nayak:2005un}.

The second-generation TDI considers the most general case that the six armlengths are time dependent, for which the operators do not commute and form a noncommutative polynomial ring.
In this case, the algorithm to compute the analogous Gröbner basis for the
noncommutative case did not terminate.
However, slowly changing armlengths permit the problem a perturbation over the static case.
The algorithm based on geometric TDI \cite{Vallisneri:2005ji} identified a significantly large number of second-generation TDI combinations \cite{Muratore:2020mdf,Muratore:2021uqj}, but could neither verify their independence nor determine the dimensionality of the second‑generation TDI space.
An analytic technique, known as the lifting operation, constructs higher-order TDI combinations by lifting up the generators of the first-generation TDI space,
and can exactly cancel the laser noise when the delays are characterized by small interspacecraft velocities \cite{Tinto:2022zmf}.

The sensitivity curve of a TDI combination, defined as the ratio between its noise power spectral density (PSD) and the sky‑ and polarization‑averaged response function \cite{Larson:1999we,Robson:2018ifk}, characterizes the overall performance of the detector.
For higher‑order TDI combinations, the averaged response function of space-based GW detectors becomes increasingly complex.
Analytical expressions and accurate approximations for the averaged response functions are therefore useful for constructing sensitivity curves efficiently, comparing different TDI observables, identifying the frequency dependence of the detector response, and providing benchmarks for numerical calculations \cite{Zhang:2021kkh,Zhang:2021wwd}.
They are particularly convenient in exploratory studies where detector parameters, noise assumptions, nominal arm lengths, or TDI configurations are varied repeatedly, for example in mission-design studies, sensitivity forecasts, and rapid comparisons of different interferometric observables.
The analytical averaged response functions for the first-generation Michelson combinations \cite{Zhang:2020khm,Zhang:2019oet} and their corresponding orthogonal $A, E, T$ channels \cite{Wang:2021jsv} have been derived previously.

At the same time, the sky- and polarization-averaged response has a limited scope of applicability: it is not intended to replace the full detector response in parameter estimation of deterministic sources, where the source location and polarization are parameters to be estimated.
Similarly, for a stochastic gravitational-wave background with a fixed detector orbit and a fixed TDI configuration, the corresponding transfer function can be precomputed and does not need to be recalculated at every step of parameter inference.

In this work, we derive analytical expressions and high-accuracy approximate formulas of averaged response functions for the second-generation TDI combinations constructed using the lifting operation, as well as for the orthogonal $A, E, T$ channels formed from these combinations.
The closed-form derivations are performed under the instantaneous equal-arm, equilateral triangular-constellation approximation.
Within the stationary phase approximation (SPA) treatment, the common armlength $L$ entering the expressions can be understood as the armlength evaluated at the stationary phase time, $L(t_f)$.
Therefore, the results should be understood as equal-arm analytical benchmarks; corrections due to unequal armlengths among different arms are not included in the present closed-form formulas.

The paper is organized as follows. 
In Sec. \ref{sec:R}, we derive the sky‑ and polarization‑averaged response function for a general TDI combination.
In Sec. \ref{sec:sn}, we introduce an inverse light‑path operator $\mathcal{P}_{i_{1}i_{2}i_{3}\ldots i_{n-1}i_{n}}$, which allows second-generation TDI combinations to be represented in a simple and compact form and provides a concise description of light propagation.
We then derive the analytical averaged response functions for the second-generation TDI Michelson, Sagnac-type ($\alpha, \beta, \gamma$), Monitor, Beacon, Relay, and fully symmetric Sagnac ($\zeta$) combinations, as well as for the orthogonal $A, E, T$ channels formed from them.
Simple and high‑accuracy approximate formulas are also provided.
Finally, we present the noise PSDs and the corresponding sensitivity curves of these combinations and their corresponding orthogonal $A, E, T$ channels.
The paper concludes in Sec. \ref{sec:conclusion}.
The validity of the analytical expressions is confirmed in Appendix \ref{app:num} by numerical results obtained via Monte Carlo integration.

\section{Averaged response functions}
\label{sec:R}

\subsection{Polarization tensor}

In the spherical coordinate system $(r,\theta,\phi)$, we introduce an orthonormal triad $(\boldsymbol{\hat{\theta}},\,\boldsymbol{\hat{\phi}},\,\boldsymbol{\hat{k}})$ to describe the polarization basis and propagation direction of GW.
The propagation direction $\boldsymbol{\hat{k}}$ naturally satisfies $\boldsymbol{\hat{k}} = \boldsymbol{\hat{\theta}} \times \boldsymbol{\hat{\phi}}$.
In Cartesian components, the three unit vectors are given by  
\begin{equation}
\label{eq:basis}
\begin{split}
\boldsymbol{\hat{\theta}}=&(\cos\theta\cos\phi, \cos\theta\sin\phi, -\sin\theta),\\
\boldsymbol{\hat{\phi}}=&(-\sin\phi,\cos\phi,0),\\
\boldsymbol{\hat{k}}=&(\sin\theta\cos\phi,\sin\theta\sin\phi,\cos\theta).
\end{split}
\end{equation}
Based on this basis, we can construct a complete set of symmetric polarization tensors $\boldsymbol{e}^{\mathcal{A}}$ characterizing different GW modes $\mathcal{A}$.  
For the two transverse–traceless tensor modes (``+" and ``$\times$"), the polarization tensors are defined as 
\begin{equation}
\label{eq:eij}
\begin{aligned}
\boldsymbol{e}^{+}_{ij} =& \boldsymbol{\hat{\theta}}_i\boldsymbol{\hat{\theta}}_j
- \boldsymbol{\hat{\phi}}_i\boldsymbol{\hat{\phi}}_j, 
& \boldsymbol{e}^{\times}_{ij} =& \boldsymbol{\hat{\theta}}_i\boldsymbol{\hat{\phi}}_j
+\boldsymbol{\hat{\phi}}_i\boldsymbol{\hat{\theta}}_j.
\end{aligned}
\end{equation}
The polarization angle $\psi$ describes a rotation of the polarization axes around the propagation direction $\boldsymbol{\hat{k}}$, leading to a rotated basis 
\begin{equation}
\label{eq:pq}
\begin{split}
\boldsymbol{\hat{p}}=\cos\psi\boldsymbol{\hat{\theta}}+\sin\psi\boldsymbol{\hat{\phi}},\quad
\boldsymbol{\hat{q}}=-\sin\psi\boldsymbol{\hat{\theta}}+\cos\psi\boldsymbol{\hat{\phi}}.
\end{split}
\end{equation}
The corresponding polarization tensors in the rotated basis are then
\begin{equation}
\label{eq:epsilon}
\begin{aligned}
\boldsymbol{\epsilon}^{+}_{ij} = \boldsymbol{e}^{+}_{ij}\cos 2\psi+ \boldsymbol{e}^{\times}_{ij}\sin2\psi, \quad
\boldsymbol{\epsilon}^{\times}_{ij} = -\boldsymbol{e}^{+}_{ij}\sin 2\psi+\boldsymbol{e}^{\times}_{ij}\cos2\psi.
\end{aligned}
\end{equation}

\subsection{GW signal from the TDI combination}

\begin{figure}[htp]
\centering
\includegraphics[width=0.35\columnwidth]{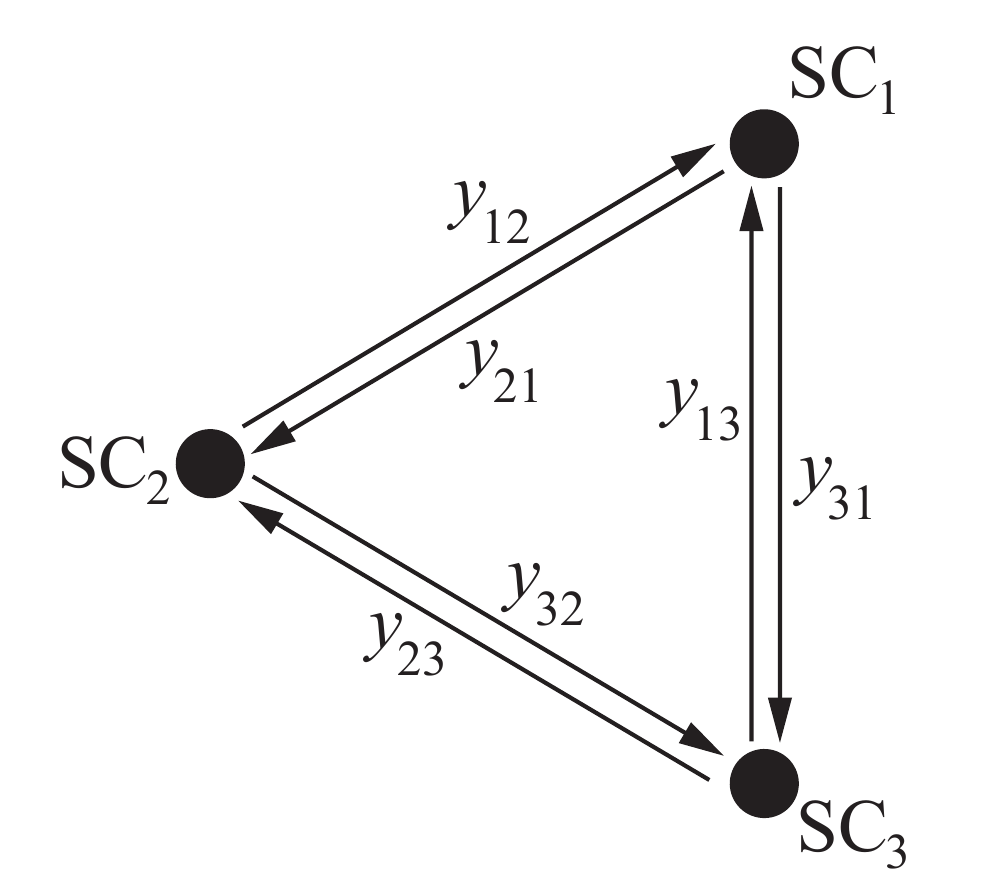}
\caption{Spacecraft configuration and definition of the one-way frequency measurements $y_{ij}$.}
\label{fig:measurements}
\end{figure}

TDI combinations are constructed as linear combinations of the six time-delayed relative frequency measurements, $y_{ij}$, obtained as light travels from spacecraft $SC_{j}$ along arm $L_{ij}$ to $SC_{i}$, where $\vec{\boldsymbol{x}}_{i}(t)$ is the position vector of spacecraft $SC_i$ in light-travel-time units, and $L_{ij}(t)=|\vec{\boldsymbol{x}}_{i}(t)-\vec{\boldsymbol{x}}_{j}(t)|$ is the corresponding one-way light-travel time.
The unit vector $\boldsymbol{\hat{n}}_{ij}$, directed from $SC_{j}$ to $SC_{i}$, denotes the orientation of arm $L_{ij}$.
The overall spacecraft configuration and the definition of the one-way frequency measurements $y_{ij}$ are illustrated in Fig. \ref{fig:measurements}.

The relative frequency shift induced by GW is given by \cite{Cornish:2002rt}
\begin{equation}
\label{eq:yij}
\begin{split}
y_{ij}(t)=& \frac{\boldsymbol{\hat{n}}_{ij}^{\text{T}}(t)\,\boldsymbol{\epsilon}^{\mathcal{A}}\,\boldsymbol{\hat{n}}_{ij}(t)}{2(1-\boldsymbol{\hat{n}}_{ij}(t)\cdot \boldsymbol{\hat{k}})} \Big[h_{\mathcal{A}}(t- L_{ij}(t)- \vec{\boldsymbol{x}}_{j}(t)\cdot\boldsymbol{\hat{k}})
-h_{\mathcal{A}}(t- \vec{\boldsymbol{x}}_{i}(t)\cdot\boldsymbol{\hat{k}})\Big].
\end{split}
\end{equation}
The delay operators are defined as
\begin{equation}
\label{eq:Dij}
\begin{split}
\mathcal{D}_{ij}x(t) =& x(t-L_{ij}(t)),\\
\mathcal{D}_{i_{1}i_{2}i_{3}\ldots i_{n-1}i_{n}} =& \mathcal{D}_{i_{1}i_{2}}\mathcal{D}_{i_{2}i_{3}}\mathcal{D}_{i_{3}i_{4}}\ldots\mathcal{D}_{i_{n-1}i_{n}}.
\end{split}
\end{equation}
Since the light-travel time $L_{ij}(t)$ varies slowly with time, we approximate chained delays as simple sums of delays rather than nested delays, i.e.,
\begin{equation}
\label{eq:Dijxt}
\begin{split}
\mathcal{D}_{i_{1}i_{2}i_{3}\ldots i_{n-1}i_{n}}x(t) =& x\left(t-\sum\limits_{k=1}^{n-1}L_{i_{k}i_{k+1}}(t)\right).
\end{split}
\end{equation}
Nested delays are considered only when cancelling the laser noise, 
whereas for the GW response and non‑laser noise we directly adopt the simple‑sum form.

The time‑domain GW signal from any TDI combination can be expressed as
\begin{equation}
\label{eq:st}
\begin{split}
s(t) = c_{12}y_{12}(t) +c_{23}y_{23}(t) + c_{31}y_{31}(t) +c_{21}y_{21}(t) + c_{32}y_{32}(t) + c_{13}y_{13}(t),
\end{split}
\end{equation}
where $c_{ij}$ are polynomials of the delay operators ($\mathcal{D}$), acting as the link coefficients of the linear combination that specifies a particular TDI combination.

The frequency-domain GW signal $s(f)$ is obtained from the time‑domain signal $s(t)$ through a Fourier transform:
\begin{equation}
\label{eq:fft}
s(f) = \int_{-\infty}^{+\infty} s(t)e^{-2 \pi i ft}dt.
\end{equation}

For heliocentric detectors such as LISA, the rotation rate of $\boldsymbol{\hat{n}}_{ij}(t)$ is only $\frac{2\pi}{\text{Year}}=2\times10^{-7}~\mathrm{Hz}$.
Therefore, for an individual harmonic of a GW waveform, the detector geometry can be evaluated at the corresponding stationary phase time $t_f$.
Under this adiabatic SPA treatment, the one-way measurement in the frequency domain can be written as
\begin{equation}
\label{eq:yijf}
\begin{split}
y_{ij}(f) =& \frac{\xi^{\mathcal{A}}_{ij}e^{-2\pi if(L_{ij}+\vec{\boldsymbol{x}}_{j}\cdot\boldsymbol{\hat{k}})}}{2(1-\mu_{ij})} \Big[1-e^{2\pi ifL_{ij}(1-\mu_{ij})}\Big]h_{\mathcal{A}}(f),
\end{split}
\end{equation}
where $\xi^{\mathcal{A}}_{ij}(t_{f})=\boldsymbol{\hat{n}}_{ij}^{\text{T}}(t_{f})\boldsymbol{\epsilon}^{\mathcal{A}}\boldsymbol{\hat{n}}_{ij}(t_{f})$ is the arm scalar, $\mu_{ij}(t_{f})=\boldsymbol{\hat{n}}_{ij}(t_{f})\cdot\boldsymbol{\hat{k}}$ is the direction cosine, and $t_{f}$ is the stationary phase time corresponding to frequency $f$ for the considered harmonic.

The frequency-domain GW signal $s(f)$ is then approximated as
\begin{equation}
\label{eq:sf}
\begin{split}
s(f) =& F^{+}(f,\theta,\phi, \psi)h_{+}(f)+ F^{\times}(f,\theta,\phi, \psi)h_{\times}(f),
\end{split}
\end{equation}
where $F^{\mathcal{A}}$ is the response function.
In the following analytical derivation, we specialize to the approximation of an instantaneous equilateral triangular constellation.
At the stationary phase time, the six optical links are assumed to have a common armlength,
\begin{equation}
	L_{ij}(t_f)=L(t_f),
\end{equation}
and the arm-opening angle is $\pi/3$.
For notational simplicity, we write $L(t_f)$ simply as $L$ in the following formulas.
Thus, the response function becomes
\begin{equation}
\label{eq:FA}
\begin{split}
&F_{A}(f,\theta,\phi,\psi)e^{2\pi i f \left( L + \vec{x}_1 \cdot \hat{k} \right)}\\
=&  c_{12}(f) \frac{\xi^A_{12}e^{-2\pi i f L\mu_{21}}}{2(1-\mu_{12})}
\Big[1 - e^{2\pi i f L (1-\mu_{12})}\Big]
+ c_{23}(f)  \frac{\xi^A_{23}e^{-2\pi i f L\mu_{31}}}{2(1-\mu_{23})}
\Big[1 - e^{2\pi i f L (1-\mu_{23})}\Big]
\\
&+ c_{31}(f)  \frac{\xi^A_{31}}{2(1-\mu_{31})}
\Big[1 - e^{2\pi i f L (1-\mu_{31})}\Big]
+ c_{21}(f) \frac{\xi^A_{21}}{2(1-\mu_{21})}
\Big[1 - e^{2\pi i f L (1-\mu_{21})}\Big]\\
&+ c_{32}(f) \frac{\xi^A_{32}e^{-2\pi i f L\mu_{21}}}{2(1-\mu_{32})}
\Big[1 - e^{2\pi i f L (1-\mu_{32})}\Big]
+ c_{13}(f) \frac{\xi^A_{13}e^{-2\pi i f L\mu_{31}}}{2(1-\mu_{13})}
\Big[1 - e^{2\pi i f L (1-\mu_{13})}\Big].
\end{split}
\end{equation}
Under the SPA, the Fourier transform of a delayed function introduces an approximate phase factor,
\begin{equation}
	\label{eq:fftdelay}
	\begin{split}
		\int_{-\infty}^{+\infty} \mathcal{D}_{ij}x(t) e^{-2 \pi i f t} d t 
		= \int_{-\infty}^{+\infty} x\big(t-L_{ij}(t)\big) e^{-2 \pi i ft} d t 
		\simeq e^{-2 \pi i f L_{ij}(t_f)} \tilde{x}(f).
	\end{split}
\end{equation}
Therefore, in the instantaneous equilateral-constellation approximation, where $L_{ij}(t_f)=L(t_f)$ for all optical links, the coefficients $c_{ij}(f)$ can be obtained by replacing each delay operator $\mathcal{D}$ with the phase factor $e^{-2 \pi i f L(t_f)}$; for brevity, this phase factor is denoted by $e^{-2\pi i fL}$ below.

In Eq. \eqref{eq:yijf}, all slowly varying geometrical quantities, including $L_{ij}$, $\vec{\boldsymbol{x}}_j$, $\xi^{\mathcal A}_{ij}$, and $\mu_{ij}$, are evaluated at the stationary phase time $t_f$ of the considered harmonic; their explicit $t_f$ dependence is suppressed for compactness.
For a waveform containing higher-order modes or eccentricity harmonics, the full frequency-domain signal can be constructed by applying Eq. \eqref{eq:yijf} to each component separately, with $t_f$ replaced by the corresponding component-dependent stationary phase time, e.g., $t_f^{\ell m}$ for the $(\ell,m)$ mode \cite{Marsat:2020rtl}.

However, after coherently summing over different modes or harmonics, the squared signal $|s(f)|^2$ generally contains cross terms between different components.
These cross terms are waveform dependent and do not, in general, vanish under the sky- and polarization-average alone.
Therefore, in the present work we restrict the analytical averaged-response derivation in Eqs. \eqref{eq:yijf}, \eqref{eq:sf}, and \eqref{eq:FA} to one harmonic of a slowly evolving waveform.
The resulting averaged response functions should be regarded as detector-level building blocks for a mode-by-mode or harmonic-by-harmonic construction of a full deterministic waveform response.

\subsection{Sensitivity curve}

The amplitude signal-to-noise ratio for a deterministic signal $s(f)$ is defined as \cite{Robson:2018ifk}
\begin{equation}
\label{eq:rho}
\begin{split}
\rho^{2} =& 4 \int_{0}^{\infty} \frac{|s(f)|^2}{P_n(f)} df,
\end{split}
\end{equation}
where $P_{n}(f)$ is the one-sided noise PSD.

The sky‑ and polarization‑averaged spectral power of the GW signal is defined as
\begin{equation}
\label{eq:average}
\begin{split}
\langle|s(f)|^{2}\rangle =& \frac{1}{8\pi^{2}}\int_{0}^{\pi}\int_{0}^{2\pi}\int_{0}^{2\pi}|s(f)|^{2}\sin \theta d \theta d \phi d \psi.
\end{split}
\end{equation}
From Eq. \eqref{eq:epsilon}, integration over the polarization angle yields
\begin{equation}
\label{eq:Fpc_relation}
\begin{split}
\frac{1}{2\pi} \int_{0}^{2\pi}F^{+}F^{\times*}d \psi = 0,\quad
\langle |F^{+}|^{2} \rangle = \langle |F^{\times}|^{2} \rangle.
\end{split}
\end{equation}
Consequently,
\begin{equation}
\label{eq:sf_average}
\begin{split}
\langle|s(f)|^{2}\rangle =& \langle|F^{+}|^{2} \rangle|h_{+}(f)|^{2}+\langle|F^{\times}|^{2} \rangle|h_{\times}(f)|^{2}\\
=& R^{+} |h_{+}(f)|^{2}+ R^{\times} |h_{\times}(f)|^{2}\\
=&\sum\limits_{\mathcal{A}=+,\times}R^{\mathcal{A}}|h_{\mathcal{A}}(f)|^{2},
\end{split}
\end{equation}
where 
\begin{equation}
\label{eq:ra}
\begin{split}
R^{\mathcal{A}}(f) =& \langle |F^{\mathcal{A}}(f,\theta,\phi,\psi)|^{2} \rangle
\end{split}
\end{equation}
is the averaged response function.

The corresponding averaged SNR becomes
\begin{equation}
\label{eq:rho_average}
\begin{split}
\langle \rho^{2} \rangle = 4 \int_{0}^{\infty} \frac{\langle|s(f)|^{2}\rangle}{P_n(f)} df
= 4 \int_{0}^{\infty} \sum\limits_{\mathcal{A}=+,\times}\frac{|h_{\mathcal{A}}(f)|^{2}}{S^{\mathcal{A}}_{n}(f)} df,
\end{split}
\end{equation}
where 
\begin{equation}
\label{eq:sn}
\begin{split}
S^{\mathcal{A}}_{n}(f) =& \frac{P_{n}(f)}{R^{\mathcal{A}}(f)}
\end{split}
\end{equation}
represents the equivalent strain noise PSD, i.e., the sensitivity curve for polarization mode $\mathcal{A}$.
From Eq. \eqref{eq:Fpc_relation}, it follows that
\begin{equation}
\label{}
\begin{split}
R^{+}(f) =& R^{\times}(f) \equiv R^{+|\times}(f),\\
S^{+}_{n}(f)=&S^{\times}_{n}(f) \equiv S^{+|\times}_{n}(f).
\end{split}
\end{equation}
Introducing the composite tensor‑mode averaged response function
\begin{equation}
\label{eq:RT}
\begin{split}
R^{\text{Tensor}}(f) = R^{+}(f) + R^{\times}(f)=2R^{+|\times}(f),
\end{split}
\end{equation} 
we obtain the composite  tensor‑mode sensitivity curve
\begin{equation}
\label{eq:sn_average_tensor}
\begin{split}
S_{n}^{\text{Tensor}}(f) =& \frac{P_{n}(f)}{R^{\text{Tensor}}(f)} = \frac{1}{2}S^{+|\times}_{n}(f),
\end{split}
\end{equation}
and the averaged SNR can be written as 
\begin{equation}
\label{eq:sf_average_tensor}
\begin{split}
\langle \rho^{2} \rangle =& 4 \int_{0}^{\infty}\frac{\frac{1}{2}\big(|h_{+}(f)|^{2}+|h_{\times}(f)|^{2}\big)}{S^{\text{Tensor}}_{n}(f)} df.
\end{split}
\end{equation}

\subsection{Averaged response function}
For the instantaneous equilateral triangular constellation considered here, the common armlength is $L(t_f)$, denoted simply by $L$ below, and the arm-opening angle is $\gamma=\pi/3$.
Owing to the rotational symmetry of the sky averaging, the constellation can be placed in the $x$--$y$ plane, with the spacecraft positioned at
\begin{equation}
\label{eq:x_i}
\begin{split}
\vec{\boldsymbol{x}}_{1} = L\left(\frac{\sqrt{3}}{2}, \frac{1}{2}, 0\right),\quad
\vec{\boldsymbol{x}}_{2} = 	\vec{\boldsymbol{0}},\quad
\vec{\boldsymbol{x}}_{3} = L\left(\frac{\sqrt{3}}{2}, -\frac{1}{2}, 0\right).
\end{split}
\end{equation}
For this configuration, the unit vectors along the detector arms are
\begin{equation}
\label{eq:nij}
\begin{split}
\boldsymbol{\hat{n}}_{12} = \left(\frac{\sqrt{3}}{2}, \frac{1}{2}, 0\right),\quad
\boldsymbol{\hat{n}}_{23} = \left(-\frac{\sqrt{3}}{2}, \frac{1}{2}, 0\right),\quad
\boldsymbol{\hat{n}}_{31} = \left(0, -1, 0\right),
\end{split}
\end{equation}
and the arm opening angle is $\gamma=\frac{\pi}{3}$.
The arm scalars can be explicitly written as
\begin{equation}
\label{eq:xi}
\begin{split}
\xi_{12}^{+} =&   \Big[ \cos^2\theta \cos^2\left(\phi-\frac{\gamma}{2}\right) - \sin^2\left(\phi-\frac{\gamma}{2}\right) \Big] \cos(2\psi)- \cos\theta \sin(2\phi-\gamma) \sin(2\psi), \\
\xi_{12}^{\times} =&  -\Big[ \cos^2\theta \cos^2\left(\phi-\frac{\gamma}{2}\right) - \sin^2\left(\phi-\frac{\gamma}{2}\right)  \Big]\sin(2\psi) - \cos\theta \sin(2\phi-\gamma) \cos(2\psi), \\
\xi_{23}^{+} =& \Big[ \cos^2\theta \cos^2\left(\phi+\frac{\gamma}{2}\right) - \sin^2\left(\phi+\frac{\gamma}{2}\right)  \Big]\cos(2\psi)- \cos\theta \sin(2\phi+\gamma) \sin(2\psi),  \\
\xi_{23}^{\times} =&  -\Big[ \cos^2\theta \cos^2\left(\phi+\frac{\gamma}{2}\right) - \sin^2\left(\phi+\frac{\gamma}{2}\right) \Big]\sin(2\psi) - \cos\theta \sin(2\phi+\gamma) \cos(2\psi),  \\
\xi_{31}^{+} =& 4\sin^{2}\left(\frac{\gamma}{2}\right) \Big[ \left( \cos^2\theta \sin^2\phi - \cos^2\phi \right) \cos(2\psi) + \cos\theta \sin(2\phi) \sin(2\psi)  \Big], \\
\xi_{31}^{\times} =& 4\sin^2\left(\frac{\gamma}{2}\right) \Big[ -\left( \cos^2\theta \sin^2\phi - \cos^2\phi \right) \sin(2\psi) + \cos\theta \sin(2\phi) \cos(2\psi)  \Big],
\end{split}
\end{equation}
and the direction cosines between the arm and the GW propagation direction are
\begin{equation}
\label{}
\begin{split}
\mu_{12} = \sin \theta \cos\left(\phi- \frac{\gamma}{2}\right),\quad
\mu_{23} = -\sin \theta \cos\left(\phi+ \frac{\gamma}{2}\right),\quad
\mu_{31} = -\sin \theta \sin \phi.
\end{split}
\end{equation}

\begin{figure}[htp]
\centering
\includegraphics[width=0.7\columnwidth]{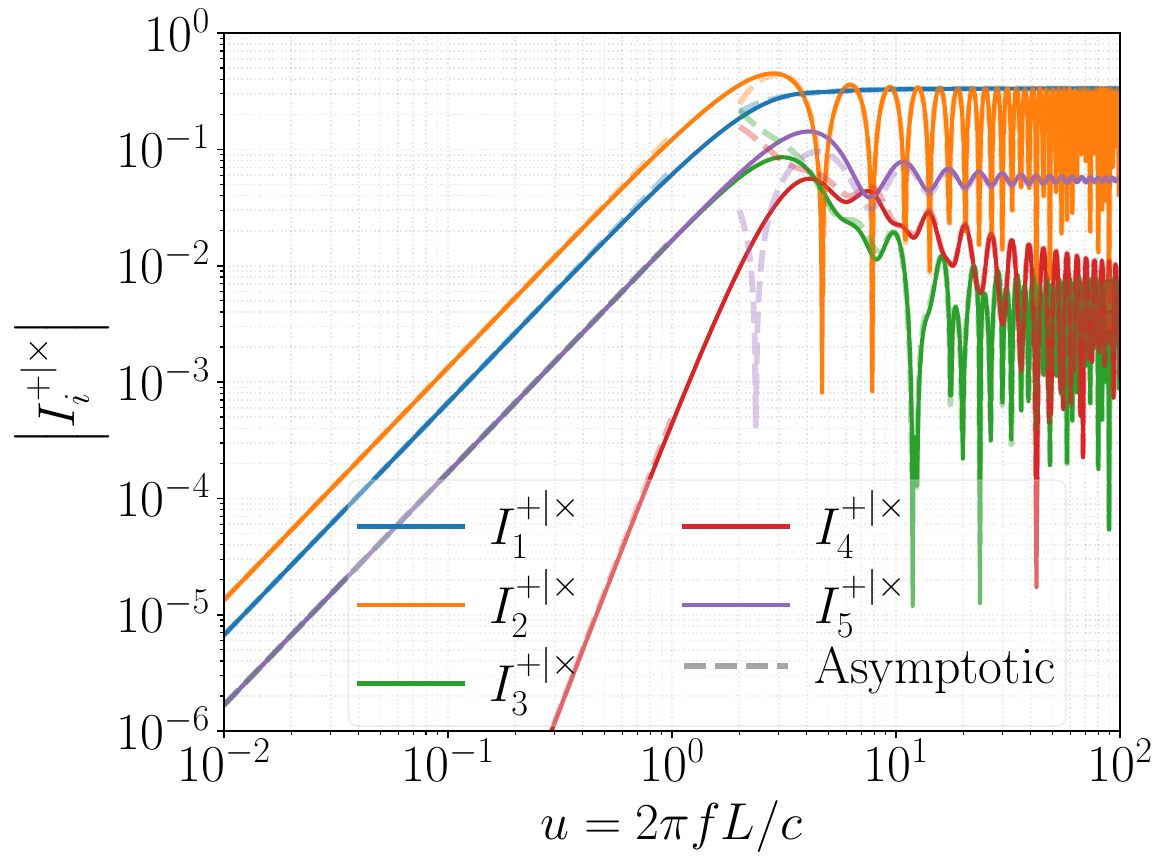}
\caption{Five basis sky- and polarization-averaged response integrals
$\left|I^{+|\times}_{i}\right|$ as functions of the dimensionless frequency
$u = 2\pi fL$.
The solid curves represent the exact numerical evaluations, 
while the dashed lines show the low‑ and high‑frequency asymptotic approximations listed in Table \ref{tab:asymptotics}.
}
\label{fig:BaseIntegrals}
\end{figure}

Using the cyclic permutation symmetry $(1\to 2\to 3\to 1)$, the sky- and polarization-averaged response function can be written as
\begin{equation}
\label{eq:Ra}
\begin{split}
R^{\mathcal{A}} =& C_{1}I^{\mathcal{A}}_{1}+C_{2}I^{\mathcal{A}}_{2}+C_{3}I^{\mathcal{A}}_{3}+C_{4}I^{\mathcal{A}}_{4}+C_{5}I^{\mathcal{A}}_{5},
\end{split}
\end{equation}
where the composite coefficients $C_i$ are algebraic combinations of the TDI link coefficients $c_{ij}$:
\begin{equation}
\label{eq:C_i}
\begin{split}
C_{1}=&|c_{12}|^{2}+|c_{23}|^{2}+|c_{31}|^{2}+|c_{21}|^{2}+|c_{32}|^{2}+|c_{13}|^{2},
\\
C_{2}=&\Re \Big[c_{12}c_{21}^{*}+c_{23}c_{32}^{*}+c_{31}c_{13}^{*}\Big],\\
C_{3}=&\Re\Big[c_{12}c_{23}^{*}+c_{23}c_{31}^{*}+c_{31}c_{12}^{*}+c_{21}c_{32}^{*}+c_{32}c_{13}^{*}+c_{13}c_{21}^{*}\Big],\\
C_{4}=&\Im\Big[c_{12}c_{23}^{*}+c_{23}c_{31}^{*}+c_{31}c_{12}^{*}-c_{21}c_{32}^{*}-c_{32}c_{13}^{*}-c_{13}c_{21}^{*}\Big], \\
C_{5}=&\Re\Big[c_{12}c_{32}^{*}+c_{23}c_{13}^{*}+c_{31}c_{21}^{*}+c_{21}c_{23}^{*}+c_{32}c_{31}^{*}+c_{13}c_{12}^{*}\Big],
\end{split}
\end{equation}
and the five basis sky‑ and polarization‑averaged integrals are
\begin{equation}
\label{}
\begin{split}
I^{\mathcal{A}}_{1} =& \Big\langle\Big[1 - \cos\Big(u(1-\mu_{12})\Big)\Big]\frac{(\xi^{\mathcal{A}}_{12})^{2}}{2(1-\mu_{12})^{2}}\Big\rangle,
\\
I^{\mathcal{A}}_{2} =& \Big\langle\Big[\cos(u \mu_{12})-\cos u\Big]\frac{(\xi^{\mathcal{A}}_{12})^{2}}{(1-\mu_{12})(1+\mu_{12})}\Big\rangle,\\
I^{\mathcal{A}}_{3} =&\Big\langle\Big[\cos (u \mu_{12})+\cos (u \mu_{23})-\cos u- \cos (u\mu_{31}+u)\Big]\frac{\xi^{\mathcal{A}}_{12}\xi^{\mathcal{A}}_{23}}{2(1-\mu_{12})(1-\mu_{23})}\Big\rangle,\\
I^{\mathcal{A}}_{4} =&\Big\langle\Big[\sin (u \mu_{12})+\sin (u \mu_{23})-\sin u+ \sin (u\mu_{31}+u)\Big]\frac{\xi^{\mathcal{A}}_{12}\xi^{\mathcal{A}}_{23}}{2(1-\mu_{12})(1-\mu_{23})}\Big\rangle,
\\
I^{\mathcal{A}}_{5} =&\Big\langle\Big[1+\cos (u \mu_{31})-\cos (u \mu_{12}-u)- \cos (u\mu_{23}+u)\Big]\frac{\xi^{\mathcal{A}}_{12}\xi^{\mathcal{A}}_{23}}{2(1-\mu_{12})(1+\mu_{23})}\Big\rangle,
\end{split}
\end{equation}
where $u=2\pi fL$.
The sky‑ and polarization‑averaged integrals are equal for the two polarizations ($I^{+}=I^{\times}\equiv I^{+|\times}$).
Following a procedure similar to that in Ref. \cite{Zhang:2020khm}, the resulting expressions are
\begin{equation}
\label{}
\begin{split}
I^{+|\times}_{1} =&  \frac{1}{3}-\frac{1}{2u^{2}}+\frac{\sin 2u}{4u^{3}},\\
I^{+|\times}_{2} =& \frac{\sin u}{u^{3}}-\Big[\frac{1}{3}+\frac{1}{u^{2}}\Big]\cos u,\\
I^{+|\times}_{3} =& -\frac{9}{32u^{2}}+\Big[\frac{1}{2u^{3}}+\frac{3}{4u}\Big]\sin u+\Big[\frac{3}{4}\text{Si}(u)-\frac{3}{2}\text{Si}(2u)+\frac{3}{4}\text{Si}(3u)\Big]\sin u\\
&+\Big[\frac{5}{32u^{3}}-\frac{9}{32u}\Big]\sin 2u+\Big[-\frac{5}{24}+\frac{3}{4}\ln\frac{4}{3}-\frac{1}{2u^{2}}\Big]\cos u\\
&+\Big[\frac{3}{4}\text{Ci}(u)-\frac{3}{2}\text{Ci}(2u)+\frac{3}{4}\text{Ci}(3u)\Big]\cos u-\frac{\cos 2u}{32u^{2}},\\
I^{+|\times}_{4} =& -\frac{5}{32u^{3}}-\frac{15}{32u}+\Big[-\frac{5}{24}+\frac{3}{4}\ln\frac{4}{3}+\frac{1}{4u^{2}}\Big]\sin u\\
&+\Big[\frac{3}{4}\text{Ci}(u)-\frac{3}{2}\text{Ci}(2u)+\frac{3}{4}\text{Ci}(3u)\Big]\sin u+\frac{\sin 2u}{32u^{2}}+\frac{3}{4u}\cos u\\
&+\Big[-\frac{3}{4}\text{Si}(u)+\frac{3}{2}\text{Si}(2u)-\frac{3}{4}\text{Si}(3u)\Big]\cos u+\Big[\frac{5}{32u^{3}}-\frac{9}{32u}\Big]\cos 2u,\\
I^{+|\times}_{5} =& \frac{7}{24}-\frac{\ln 2}{2}-\frac{5}{8u^{2}}
-\frac{\text{Ci}(u)}{2}+\frac{1}{2}\text{Ci}(2u)\\
&+\Big[\frac{5}{16u^{3}}+\frac{11}{16u}\Big]\sin u
+\Big[\frac{1}{4u^{3}}-\frac{1}{4u}\Big]\sin 2u-\frac{5\cos u}{16u^{2}}+\frac{\cos 2u}{8u^{2}},
\end{split}
\end{equation}
where $\text{Si}(u)$ and $\text{Ci}(u)$ denote the sine‑ and cosine‑integral functions, respectively.
The asymptotic forms of $I^{+|\times}_{i}$ are summarized in Table \ref{tab:asymptotics} and illustrated in Fig. \ref{fig:BaseIntegrals}.

\begin{table}[ht]
\centering
\renewcommand{\arraystretch}{2}
\resizebox{0.55\columnwidth}{!}{
\begin{tabular}{|c|c|c|}
	\hline
	& $u\ll1$ & $u\gg1$ \\
	\hline
	$\text{Si}(u)$ & $u-\frac{u^3}{18}$ & $\frac{\pi }{2} - \frac{\cos u}{u}$ \\
	\hline
	$\text{Ci}(u)$ & $\gamma_{E}+\ln u$ & $\frac{\sin u}{u}- \frac{\cos u}{u^{2}}$ \\
	\hline
	$I^{+|\times}_{1}$ & $\frac{u^2}{15}$ & $\frac{1}{3}-\frac{1}{2u^{2}}$ \\
	\hline
	$I^{+|\times}_{2}$ & $\frac{2 u^2}{15}$ & $-\big(\frac{1}{3}+\frac{1}{u^{2}}\big)\cos u$ \\
	\hline
	$I^{+|\times}_{3}$ & $-\frac{u^2}{60}$ & $\begin{array}{@{}c@{}}
		\big(-\frac{5}{24}+\frac{3}{4}\ln\frac{4}{3}\big)\cos u
		-\frac{\sin2u}{32u}\\[2pt]
		-\frac{1}{96u^2}(99+12\cos u+11\cos2u)
	\end{array}$\\
	\hline
	$I^{+|\times}_{4}$ & $\frac{u^5}{2016}$ & $\big(-\frac{5}{24}+\frac{3}{4} \ln \frac{4}{3}\big) \sin u+ \frac{9-\cos2u}{32u}$ \\
	\hline
	$I^{+|\times}_{5}$ & $-\frac{u^2}{60}$ & $\frac{7}{24}-\frac{\ln 2}{2}+\frac{3 \sin u}{16 u}$ \\
	\hline
\end{tabular}
}
\caption{Asymptotic expressions of the sine and cosine integrals
$\text{Si}(u)$, $\text{Ci}(u)$,
and of the sky- and polarization-averaged response integrals $I^{+|\times}_{i}$.
Here $u=2\pi fL$, $\gamma_{E}$ is the Euler constant,
and $I^{+}=I^{\times}\equiv I^{+|\times}$.}
\label{tab:asymptotics}
\end{table}

\subsection{Optimal sensitivity}

In the ideal equal-arm and identical-noise limit, the basic TDI combinations $(a,b,c)$ that are related by cyclic permutations can be linearly combined to construct the orthogonal $A,E,T$ modes \cite{Prince:2002hp},
\begin{equation}
	\label{eq:AET}
	\begin{split}
		A(a,b,c) =& \frac{1}{\sqrt{2}}(c-a),\\
		E(a,b,c) =& \frac{1}{\sqrt{6}}(a-2b+c),\\
		T(a,b,c) =& \frac{1}{\sqrt{3}}(a+b+c).
	\end{split}
\end{equation}
Here the noise independence of the $A,E,T$ channels assumes that the proof-mass and optical-path noises in different links are mutually uncorrelated, that the corresponding noise PSDs are identical for the three arms, and that the covariance matrix of $(a,b,c)$ has the cyclic symmetry implied by the instantaneous equilateral constellation.
Under these conditions, the transformation in Eq. \eqref{eq:AET} diagonalizes the noise covariance matrix, so that the $A,E,T$ channels provide mutually independent data streams.
Because the covariance matrix is diagonal in this idealized limit, the overall sensitivity and the network SNR can be obtained by direct summation of the individual contributions as
\begin{equation}
\label{eq:rhosum}
\begin{split}
\rho^{2}_{\text{opt}(a,b,c)} =&\rho^{2}_{A(a,b,c)}+\rho^{2}_{E(a,b,c)}+\rho^{2}_{T(a,b,c)}.
\end{split}
\end{equation}
Thus, for a given polarization mode $\mathcal{A}$, the optimal sensitivity is
\begin{equation}
\label{eq:snopt}
\begin{split}
&S^{\mathcal{A}}_{n;\text{opt}(a,b,c)}(f) \\
=& \Big(\big[S^{\mathcal{A}}_{n;A(a,b,c)}(f)\big]^{-1}+\big[S^{\mathcal{A}}_{n;E(a,b,c)}(f)\big]^{-1}\\
&+\big[S^{\mathcal{A}}_{n;T(a,b,c)}(f)\big]^{-1}\Big)^{-1}\\
=& \Big(2\big[S^{\mathcal{A}}_{n;A|E(a,b,c)}(f)\big]^{-1}+\big[S^{\mathcal{A}}_{n;T(a,b,c)}(f)\big]^{-1}\Big)^{-1}.
\end{split}
\end{equation}

If these ideal assumptions are relaxed, for example in an unequal-arm constellation or in the presence of unequal or correlated noise sources, the $A,E,T$ channels are not guaranteed to be exactly noise independent.
In that case, the full noise covariance matrix should be used to construct the optimal sensitivity, rather than the simple inverse-sensitivity sum in Eq. \eqref{eq:snopt}.

\section{Analytical sensitivity curves of time-delay interferometry}
\label{sec:sn}

After laser frequency noise is suppressed by the TDI technique, several secondary noise sources may still be present in the data.
In this work, we follow the standard simplified sensitivity model and include the two dominant residual instrumental noises: proof-mass noise and optical-path noise.
Other contributions, such as clock noise, EOM modulation noise, plasma noise, calibration residuals, and data-processing residuals, are assumed to be suppressed by preprocessing or subdominant, and are not included in the analytical noise model considered here.
The PSDs of the proof-mass and optical-path noises, denoted by $S_{y}^{\text{proof mass}}$ and $S_{y}^{\text{optical path}}$, are related to the position noise $S_x$ and acceleration noise $S_a$ via $S_{y}^{\text{proof mass}} = S_a/(2\pi fc)^2$ and $S_{y}^{\text{optical path}} = S_x\left(2\pi f/c\right)^2$.

For LISA, $S_x = (1.5 \times 10^{-11} \text{ m})^2$ $\text{Hz}^{-1}$, $S_a = (3 \times 10^{-15} \text{ m\,s}^{-2})^2 \Big[1+\left(\frac{0.4 \text{ mHz}}{f}\right)^2\Big]$ $\text{Hz}^{-1}$, and $L=2.5\times10^9~{\rm m}/c$  \cite{LISA:2017pwj}.
For Taiji, $S_x = (8\times10^{-12} \text{ m})^2$ $\text{Hz}^{-1}$, $S_a = (3\times10^{-15} \text{ m\,s}^{-2})^2$ $\text{Hz}^{-1}$, and $L=3\times10^9~{\rm m}/c$ \cite{Ruan:2020smc}.
For TianQin, $S_x = (10^{-12}\text{ m})^2$ $\text{Hz}^{-1}$, $S_a = (10^{-15} \ \text{ m\,s}^{-2})^2$ $\text{Hz}^{-1}$, and $L=\sqrt{3}\times10^8~{\rm m}/c$ \cite{Luo:2015ght}.
Accordingly, for LISA,
$S_{y}^{\text {proof mass}}(f)= 2.5 \times 10^{-48} \left(\frac{1 \text{ Hz}}{f}\right)^{2}
\left[1+\left(\frac{0.4 \text{ mHz}}{f}\right)^2\right]$ $\text{Hz}^{-1}$, $S_{y}^{\text {optical path}}(f) = 9.9 \times 10^{-38} \left(\frac{f}{1 \text{ Hz}}\right)^{2}$ $\text{Hz}^{-1}$;
For Taiji, $S_{y}^{\text {proof mass}}(f)= 2.5 \times 10^{-48}\left(\frac{1 \text{ Hz}}{f}\right)^{2}$ $\text{Hz}^{-1}$,
$S_{y}^{\text {optical path}}(f) =2.8 \times 10^{-38} \left(\frac{f}{1 \text{ Hz}}\right)^{2}$ $\text{Hz}^{-1}$;
For TianQin, $S_{y}^{\text {proof mass}}(f)= 2.8 \times 10^{-49} \left(\frac{1 \text{ Hz}}{f}\right)^{2}$ $\text{Hz}^{-1}$,
$S_{y}^{\text {optical path}}(f) = 4.4 \times 10^{-40} \left(\frac{f}{1 \text{ Hz}}\right)^{2}$ $\text{Hz}^{-1}$.
For convenience, we define $c^{\text{pm}}$ as the constant coefficient of $f^{-2}$ in $S_{y}^{\text{proof mass}}$  
and $c^{\text{op}}$ as that of $f^{2}$ in $S_{y}^{\text{optical path}}$.

To represent TDI combinations in a simple and compact form, we introduce an inverse light‑path operator $\mathcal{P}_{i_{1}i_{2}i_{3}\ldots i_{n-1}i_{n}}$.
The subscripts specify the light path in the reverse order of propagation; for instance, $\mathcal{P}_{123}$ represents a light path $1 \leftarrow 2 \leftarrow 3$.
It can be expanded as
\begin{equation}
\label{eq:P}
\begin{split}
\mathcal{P}_{i_{1}i_{2}i_{3}\ldots i_{n-1}i_{n}} =& y_{i_{1}i_{2}}+ \mathcal{D}_{i_{1}i_{2}}y_{i_{2}i_{3}}+ \mathcal{D}_{i_{1}i_{2}i_{3}}y_{i_{3}i_{4}}+\ldots+ \mathcal{D}_{i_{1}i_{2}i_{3}\ldots i_{n-1}}y_{i_{n-1}i_{n}}.
\end{split}
\end{equation}

\subsection{Michelson combinations (X, Y, Z)}
\begin{figure}[htp]
\centering
\includegraphics[width=0.6\columnwidth]{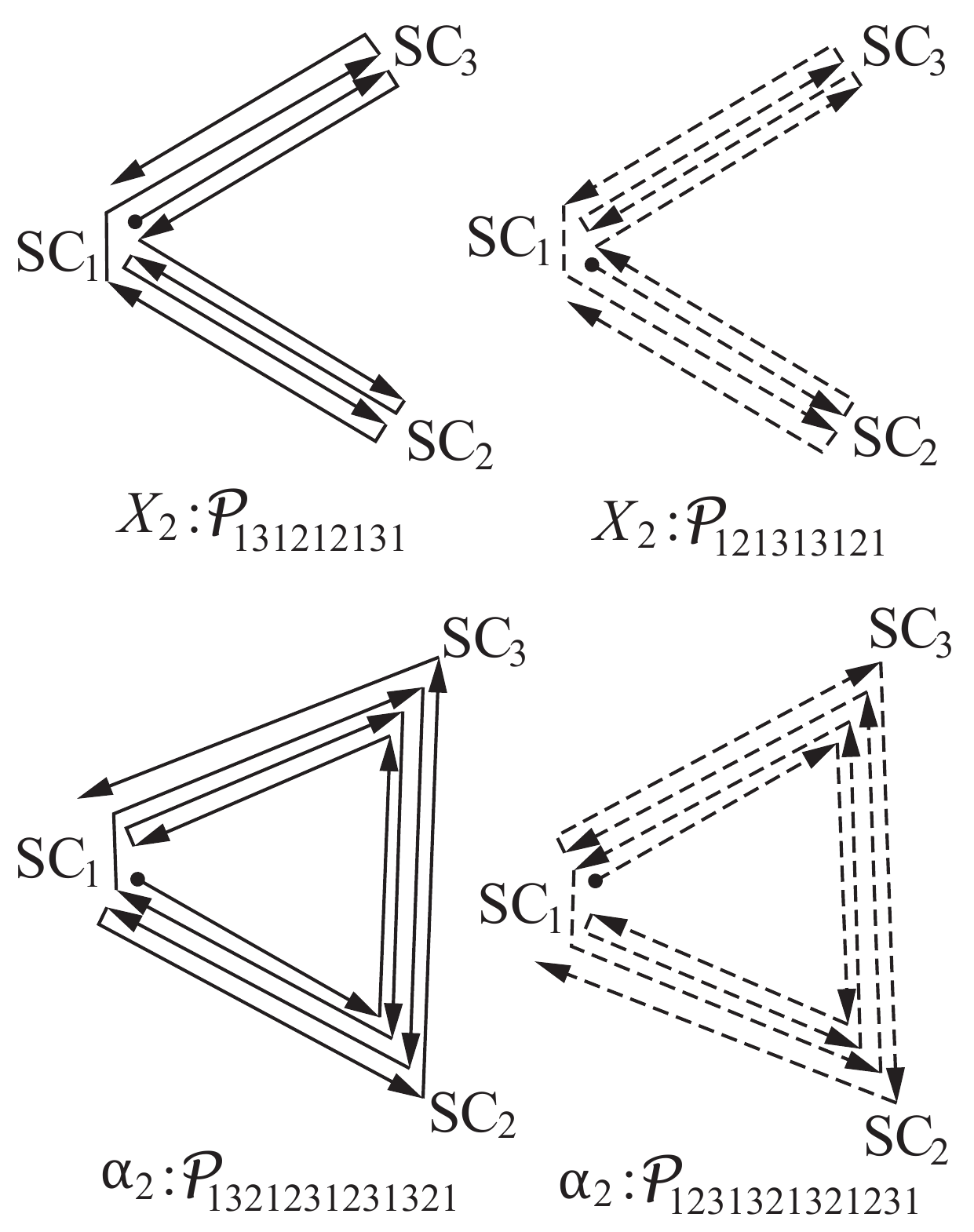}
\caption{Light paths of the second-generation TDI combinations $X_{2}$ and  $\alpha_{2}$.}
\label{fig:path}
\end{figure}

The equal-arm Michelson interferometer can be treated as the zeroth-generation Michelson combination, 
\begin{equation}
\label{eq:x0}
\begin{split}
X_{0} =& \mathcal{P}_{131} - \mathcal{P}_{121}\\
=& y_{13}+ \mathcal{D}_{13}y_{31}- [y_{12}+ \mathcal{D}_{12}y_{21}].
\end{split}
\end{equation}
The first-generation Michelson combination is defined as
\begin{equation}
\label{eq:x1}
\begin{split}
X_{1} =& \mathcal{P}_{13121} - \mathcal{P}_{12131}\\
=&\left[y_{13}+\mathcal{D}_{13} y_{31}+\mathcal{D}_{131} y_{12}+\mathcal{D}_{1312} y_{21}\right] \\
& -\left[y_{12}+\mathcal{D}_{12} y_{21}+\mathcal{D}_{121} y_{13}+\mathcal{D}_{1213} y_{31}\right].
\end{split}
\end{equation}
The second-generation Michelson combination is
\begin{equation}
\label{eq:X2}
\begin{split}
X_{2} =& \mathcal{P}_{131212131} - \mathcal{P}_{121313121}\\
= & y_{13}+\mathcal{D}_{13} y_{31}+\mathcal{D}_{131} y_{12}+\mathcal{D}_{1312} y_{21}+\mathcal{D}_{13121} y_{12}\\
&+\mathcal{D}_{131212} y_{21}+\mathcal{D}_{1312121} y_{13}  +\mathcal{D}_{13121213} y_{31}\\
&-\Big[y_{12}+\mathcal{D}_{12} y_{21}+\mathcal{D}_{121} y_{13}+\mathcal{D}_{1213} y_{31}+\mathcal{D}_{12131} y_{13}\\
&+\mathcal{D}_{121313} y_{31} +\mathcal{D}_{1213131} y_{12}+\mathcal{D}_{12131312} y_{21}\Big].
\end{split}
\end{equation}
The light paths of $X_{2}$ are illustrated in Fig. \ref{fig:path}.

The Michelson combinations $Y$ and $Z$ can be obtained by the cyclic
permutation of the indices $1\to2\to3\to1$.
For example,
\begin{equation}
\label{eq:Y2}
\begin{split}
Y_{2} =& \mathcal{P}_{212323212} - \mathcal{P}_{232121232}\\
= &\ y_{21}+\mathcal{D}_{21} y_{12}+\mathcal{D}_{212} y_{23}+\mathcal{D}_{2123} y_{32}+\mathcal{D}_{21232} y_{23}\\
&+\mathcal{D}_{212323} y_{32}+\mathcal{D}_{2123232} y_{21}  +\mathcal{D}_{21232321} y_{12}\\
&-\Big[y_{23}+\mathcal{D}_{23} y_{32}+\mathcal{D}_{232} y_{21}+\mathcal{D}_{2321} y_{12}+\mathcal{D}_{23212} y_{21}\\
&+\mathcal{D}_{232121} y_{12} +\mathcal{D}_{2321212} y_{23}+\mathcal{D}_{23212123} y_{32}\Big],
\end{split}
\end{equation}
and
\begin{equation}
\label{eq:Z2}
\begin{split}
Z_{2} =& \mathcal{P}_{323131323} - \mathcal{P}_{313232313}\\
= & y_{32}+\mathcal{D}_{32} y_{23}+\mathcal{D}_{323} y_{31}+\mathcal{D}_{3231} y_{13}+\mathcal{D}_{32313} y_{31}\\
&+\mathcal{D}_{323131} y_{13}+\mathcal{D}_{3231313} y_{32}  +\mathcal{D}_{32313132} y_{23}\\
&-\Big[y_{31}+\mathcal{D}_{31} y_{13}+\mathcal{D}_{313} y_{32}+\mathcal{D}_{3132} y_{23}+\mathcal{D}_{31323} y_{32}\\
&+\mathcal{D}_{313232} y_{23} +\mathcal{D}_{3132323} y_{31}+\mathcal{D}_{31323231} y_{13}\Big].
\end{split}
\end{equation}

\subsubsection{Sensitivity curves of X, Y, Z}

From Eqs. \eqref{eq:st} and \eqref{eq:X2}, the coefficient $c_{12}^{X_{2}} = \mathcal{D}_{131}+\mathcal{D}_{13121}-\Big[1 +\mathcal{D}_{1213131} \Big]$.
Replacing each delay operator $\mathcal{D}$ with the phase factor $e^{-iu}$ gives
$c_{12}^{X_{2}}(f)=e^{-2 i u}+e^{-4 i u}-1-e^{-6 i u}$.
From Eq. \eqref{eq:C_i}, the composite coefficients for $X_2$ are
\begin{equation}
\label{eq:CX2}
\begin{split}
C_{1}^{X_{2}} =& 64 \sin^{2} u \sin^{2} 2u, \quad C_{2}^{X_{2}} = 16 \sin u \sin^{3} 2u, \quad
C_{3}^{X_{2}} = -16 \sin u \sin^{3} 2u,\\
C_{4}^{X_{2}} =& 32 \sin^3 u \sin^{2} 2u, \quad
C_{5}^{X_{2}} = -32 \sin^{2} u \sin^{2} 2u.
\end{split}
\end{equation}
From Eq. \eqref{eq:Ra},
the sky- and polarization-averaged response functions of the Michelson combinations, together with their asymptotic limits and high-accuracy approximations, are
\begin{equation}
\label{eq:RXYZ}
\begin{split}
&\frac{R^{+|\times}_{X_{2}|Y_{2}|Z_{2}}}{16\sin^{2}u\sin^{2}2u}=\frac{R^{+|\times}_{X_{1}|Y_{1}|Z_{1}}}{4\sin^{2}u}=R^{+|\times}_{X_{0}|Y_{0}|Z_{0}}   \\
=& \frac{5}{12} + \ln 2 - \frac{1}{u^2} + \text{Ci}(u) - \text{Ci}(2u) - \Big[\frac{5}{4u^3} + \frac{7}{4u}\Big] \sin u + \Big[\frac{1}{u^3} + \frac{1}{2u}\Big] \sin 2u \\
& + \Big[-\frac{3}{2}\text{Si}(u) + 3 \text{Si}(2u) - \frac{3}{2}\text{Si}(3u)\Big] \sin 2u + \frac{5 \cos u}{4 u^2}\\
& + \Big[\frac{1}{12} - \frac{3}{2} \ln \frac{4}{3} - \frac{1}{u^2}\Big] \cos 2u + \Big[-\frac{3}{2}\text{Ci}(u) + 3 \text{Ci}(2u) - \frac{3}{2}\text{Ci}(3u)\Big] \cos 2u\\
\sim& \begin{cases} \frac{3}{5}u^{2}, & u\ll1.56,\\
	\frac{5}{12}+\ln2+\Big[\frac{1}{12}-\frac{3}{2} \ln\frac{4}{3}\Big] \cos 2 u, & u\gg1.56 \end{cases}\\
\approx& \frac{3}{5}u^{2} \left(1+ \frac{u^{2}}{1.85-0.58\cos2u}\right)^{-1}.
\end{split}
\end{equation}
Here $u=1.56$ is the response-transition frequency, obtained by equating the low- and high-frequency asymptotic expressions;
it therefore denotes the boundary between the response function’s low- and high-frequency behaviors.
Eq. \eqref{eq:RXYZ} is consistent with the result of $X_{1}$ in Ref. \cite{Zhang:2020khm}.
Numerical evaluations of Eqs. \eqref{eq:FA} and \eqref{eq:ra} using a Monte Carlo integration technique are shown in Fig. \ref{fig:RCompare} of the Appendix.
The analytical and high-accuracy approximate expressions for the averaged response functions, and the response-transition frequency are shown in Fig. \ref{fig:Rpc}.

The noise terms of the second-generation Michelson combination are analyzed in the Appendix \ref{app:noise}.
The noise PSDs of the Michelson combinations are
\begin{equation}
\label{eq:PnX}
\begin{split}
&\frac{P_{n}^{X_{2}|Y_{2}|Z_{2}}}{16\sin^{2}u\sin^{2}2u}=\frac{P_{n}^{X_{1}|Y_{1}|Z_{1}}}{4\sin^{2}u}=P_{n}^{X_{0}|Y_{0}|Z_{0}}  \\
=& 4\big(3+\cos2u\big) S_{y}^{\text{proof mass}} + 4S_{y}^{\text{optical path}}.
\end{split}
\end{equation}

To analyze the asymptotic behavior of the noise PSD in the low- and high-frequency limits, we introduce the noise-balance frequency $f_{n}$, defined as the frequency where the noise PSD is equally contributed by the proof-mass noise and the optical-path noise.
It therefore marks the boundary between the proof-mass-dominated (low-frequency) and optical-path-dominated (high-frequency) noise regimes.
For $P_{n}^{X|Y|Z}$, this definition gives $16S_{y}^{\text{proof mass}}(f_{n})=4S_{y}^{\text{optical path}}(f_{n})$.
Consequently,
\begin{equation}
\label{}
\begin{split}
&\frac{P_{n}^{X_{2}|Y_{2}|Z_{2}}}{16\sin^{2}u\sin^{2}2u}=\frac{P_{n}^{X_{1}|Y_{1}|Z_{1}}}{4\sin^{2}u}=P_{n}^{X_{0}|Y_{0}|Z_{0}}  \\
\sim& \begin{cases} 16c^{\text{pm}}f^{-2}, & f\ll f_{n}^{X|Y|Z},\\
	4c^{\text{op}}f^{2},  &  f\gg f_{n}^{X|Y|Z},
\end{cases}
\end{split}
\end{equation}
where $f_{n}^{X|Y|Z}=(3.17\times10^{-3},4.35\times10^{-3},7.10\times10^{-3})$ Hz for LISA, Taiji, and TianQin.
The corresponding PSDs are plotted in Figs. \ref{fig:PnLISA}, \ref{fig:PnTaiji} and \ref{fig:PnTianQin}.

\begin{figure*}[htp]
\centering
\includegraphics[width=0.95\textwidth]{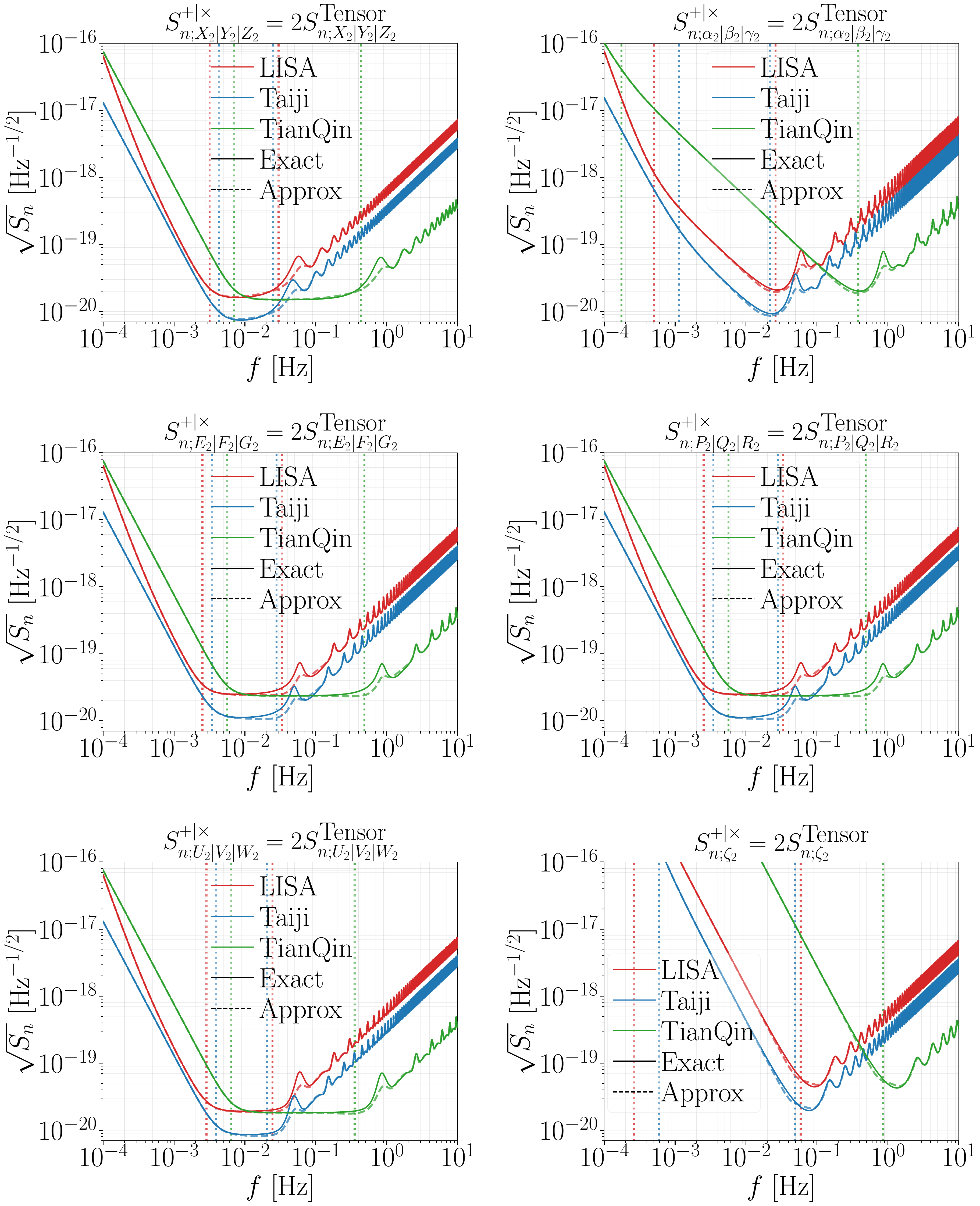}
\caption{Analytical (solid) and approximate (dashed) sensitivity curves for the second-generation TDI Michelson, Sagnac-type ($\alpha$, $\beta$, $\gamma$), Monitor, Beacon, Relay, and fully symmetric Sagnac ($\zeta$) combinations.
Results are shown for the plus, cross, and composite tensor modes. The composite tensor mode satisfies $R^{T}(f) = R^{+}(f) + R^{\times}(f) = 2R^{+|\times}(f)$.
Each solid curve is accompanied by two vertical dotted lines of the same color: the left one marks the noise-balance frequency $f_{n}$—the boundary between the proof-mass–dominated (low-frequency) and optical-path–dominated (high-frequency) noise regimes;
the right one marks the response-transition frequency—the boundary between the response function’s low- and high-frequency behaviors.
}
\label{fig:Sn}
\end{figure*}

Using Eq. \eqref{eq:sn}, the resulting sensitivity curves are
\begin{equation}
\label{}
\begin{split}
&S^{+|\times}_{n;X_{2}|Y_{2}|Z_{2}}=S^{+|\times}_{n;X_{1}|Y_{1}|Z_{1}}=S^{+|\times}_{n;X_{0}|Y_{0}|Z_{0}}\\
\equiv&S^{+|\times}_{n;X|Y|Z}\\
\sim& \begin{cases} \frac{20}{3\pi^{2}L^{2}}c^{\text{pm}}f^{-4}, & f\ll f_{n}^{X|Y|Z},\\
	\frac{5c^{\text{op}}}{3\pi^{2}L^{2}}, & f_{n}^{X|Y|Z}<f<\frac{1.56}{2\pi L},\\
	\frac{4c^{\text{op}}}{1.11-0.35\cos 2u}f^{2},  &  f\gg \frac{1.56}{2\pi L}.
\end{cases}
\end{split}
\end{equation}
Hence, the sensitivity curves of the Michelson combinations exhibit a flat section in the range $f_{n}^{X|Y|Z}<f<\frac{1.56}{2\pi L}$, as illustrated in Fig. \ref{fig:Sn}.

\subsubsection{Sensitivity curves of $A, E, T$}

\begin{figure*}[htp]
\centering
\includegraphics[width=0.95\textwidth]{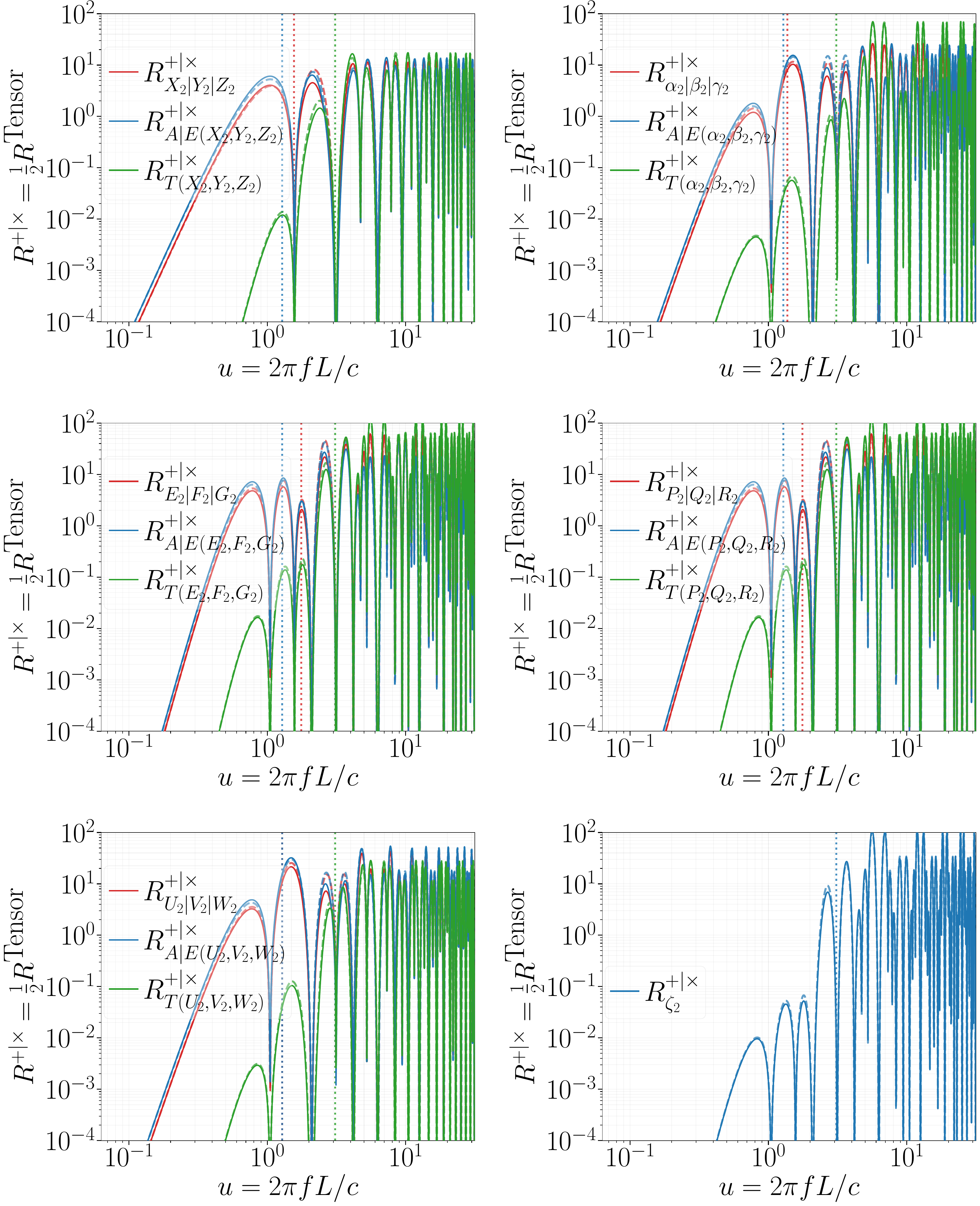}
\caption{Analytical (solid) and approximate (dashed) sky‑ and polarization‑averaged response functions of the second‑generation TDI combinations, along with their orthogonal channels. 
Each solid line is accompanied by a vertical dotted line of the same color, marking the response-transition frequency—the boundary between the response function’s low- and high-frequency behaviors.}
\label{fig:Rpc}
\end{figure*}

From Eqs. \eqref{eq:AET} and \eqref{eq:X2}, the sky- and polarization-averaged response functions of the orthogonal modes $A$ and $E$, constructed from the Michelson combinations, are
\begin{equation}
\label{eq:AETXYZ}
\begin{split}
&\frac{R^{+|\times}_{A|E(X_{2},Y_{2},Z_{2})}}{16\sin^{2}u\sin^{2}2u}=\frac{R^{+|\times}_{A|E(X_{1},Y_{1},Z_{1})}}{4\sin^{2}u}=R^{+|\times}_{A|E(X_{0},Y_{0},Z_{0})}\\
=& \frac{1}{2} + \frac{3}{4}\ln\frac{4}{3} + \frac{1}{2}\ln 2 - \frac{7}{8u^2}+ \frac{5}{4}\text{Ci}(u) - 2\text{Ci}(2u) + \frac{3}{4}\text{Ci}(3u)\\
&- \Big[\frac{35}{32u^3} + \frac{89}{32u}\Big]\sin u+ \Big[-\frac{3}{2}\text{Si}(u) + 3\text{Si}(2u) - \frac{3}{2}\text{Si}(3u)\Big]\sin u\\
&+ \Big[\frac{7}{8u^3} + \frac{7}{8u}\Big]\sin 2u+ \Big[-\frac{3}{2}\text{Si}(u) + 3\text{Si}(2u) - \frac{3}{2}\text{Si}(3u)\Big]\sin 2u\\
&+ \Big[\frac{5}{32u^3} - \frac{1}{32u}\Big]\sin 3u+ \Big[\frac{1}{6} - \frac{3}{2}\ln\frac{4}{3} + \ln 2 + \frac{25}{32u^2}\Big]\cos u\\
&+ \Big[-\frac{1}{2}\text{Ci}(u) + 2\text{Ci}(2u) - \frac{3}{2}\text{Ci}(3u)\Big]\cos u+ \Big[\frac{1}{12} - \frac{3}{2}\ln\frac{4}{3} - \frac{7}{8u^2}\Big]\cos 2u\\
&+ \Big[-\frac{3}{2}\text{Ci}(u) + 3\text{Ci}(2u) - \frac{3}{2}\text{Ci}(3u)\Big]\cos 2u- \frac{5}{32u^2}\cos 3u\\
\sim& \begin{cases} \frac{9}{10}u^{2}, & u\ll1.28,\\
	\frac{1}{2}+\frac{1}{2}\ln 2+\frac{3}{4}\ln\frac{4}{3} +\Big[\frac{1}{6}-\frac{3}{2}\ln\frac{4}{3} +\ln 2\Big]\cos u+\Big[\frac{1}{12}-\frac{3}{2}\ln\frac{4}{3}\Big]\cos 2u,
	& u\gg1.28 \end{cases}\\
\approx& \frac{9}{10}u^{2} \left(1+ \frac{u^{2}}{1.18+ 0.48\cos u-0.39 \cos2u}\right)^{-1},
\end{split}
\end{equation}
as shown in Fig. \ref{fig:Rpc}.

\begin{figure*}[htp]
\centering
\includegraphics[width=0.95\textwidth]{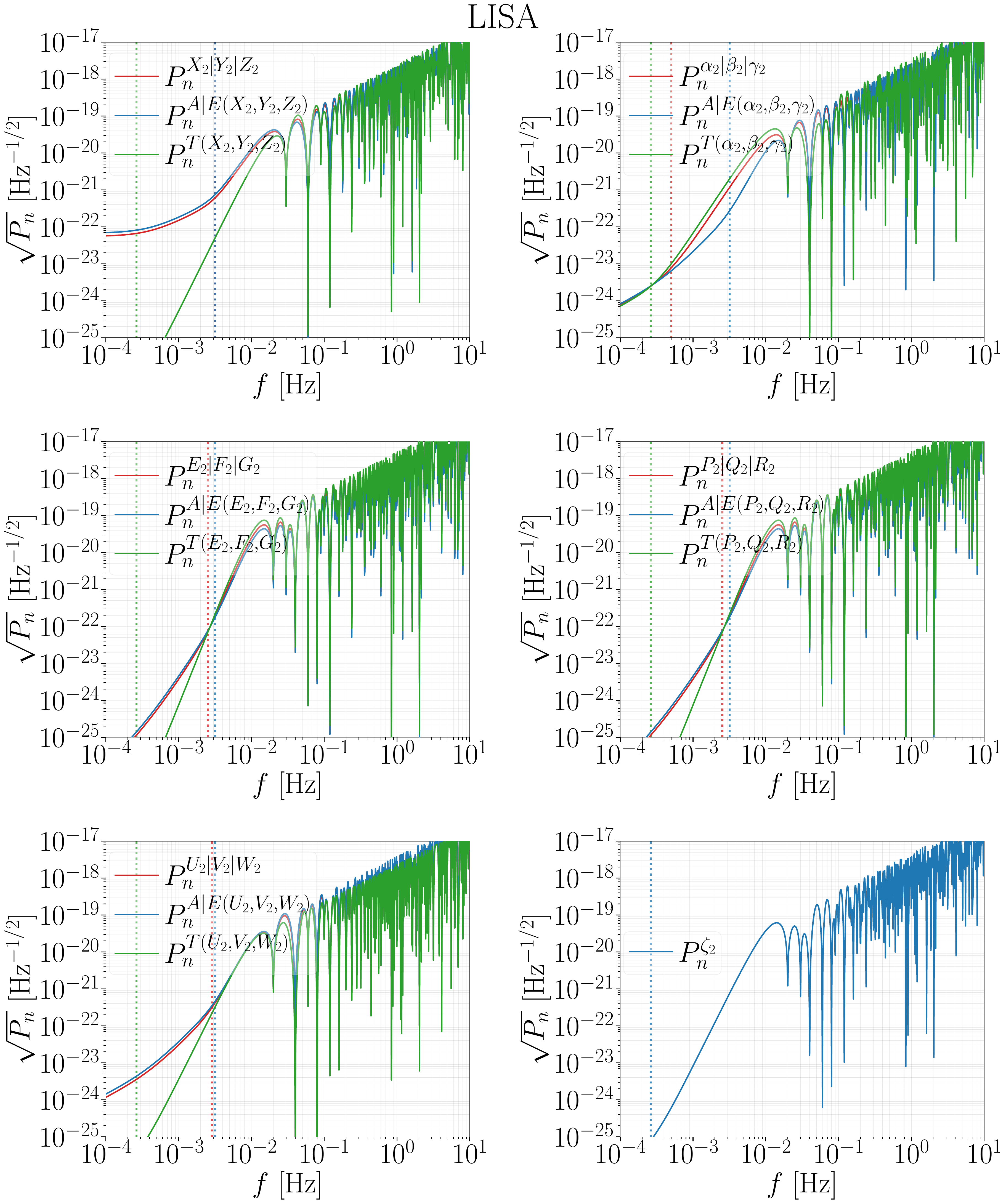}
\caption{Noise PSDs of the second‑generation TDI combinations for the LISA mission, together with the PSDs of the corresponding orthogonal combinations.
Each solid line is accompanied by a vertical dotted line of the same color, marking  the noise-balance frequency $f_{n}$—the boundary between the proof-mass–dominated (low-frequency) and optical-path–dominated (high-frequency) noise regimes.}
\label{fig:PnLISA}
\end{figure*}

\begin{figure*}[htp]
\centering
\includegraphics[width=0.95\textwidth]{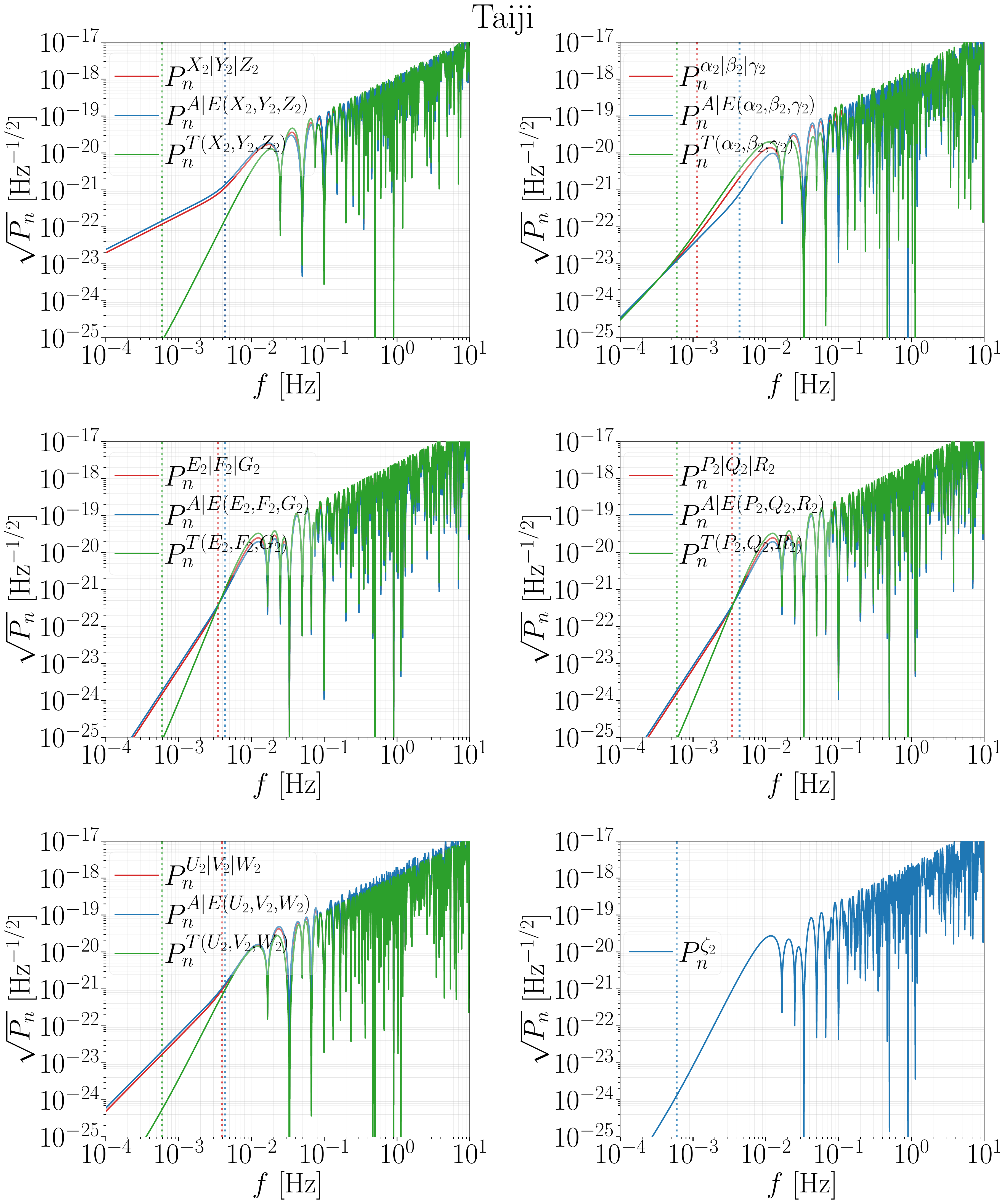}
\caption{Noise PSDs of the second‑generation TDI combinations for the Taiji mission, together with the PSDs of the corresponding orthogonal combinations.
Each solid line is accompanied by a vertical dotted line of the same color, marking  the noise-balance frequency $f_{n}$—the boundary between the proof-mass–dominated (low-frequency) and optical-path–dominated (high-frequency) noise regimes.}
\label{fig:PnTaiji}
\end{figure*}

\begin{figure*}[htp]
\centering
\includegraphics[width=0.95\textwidth]{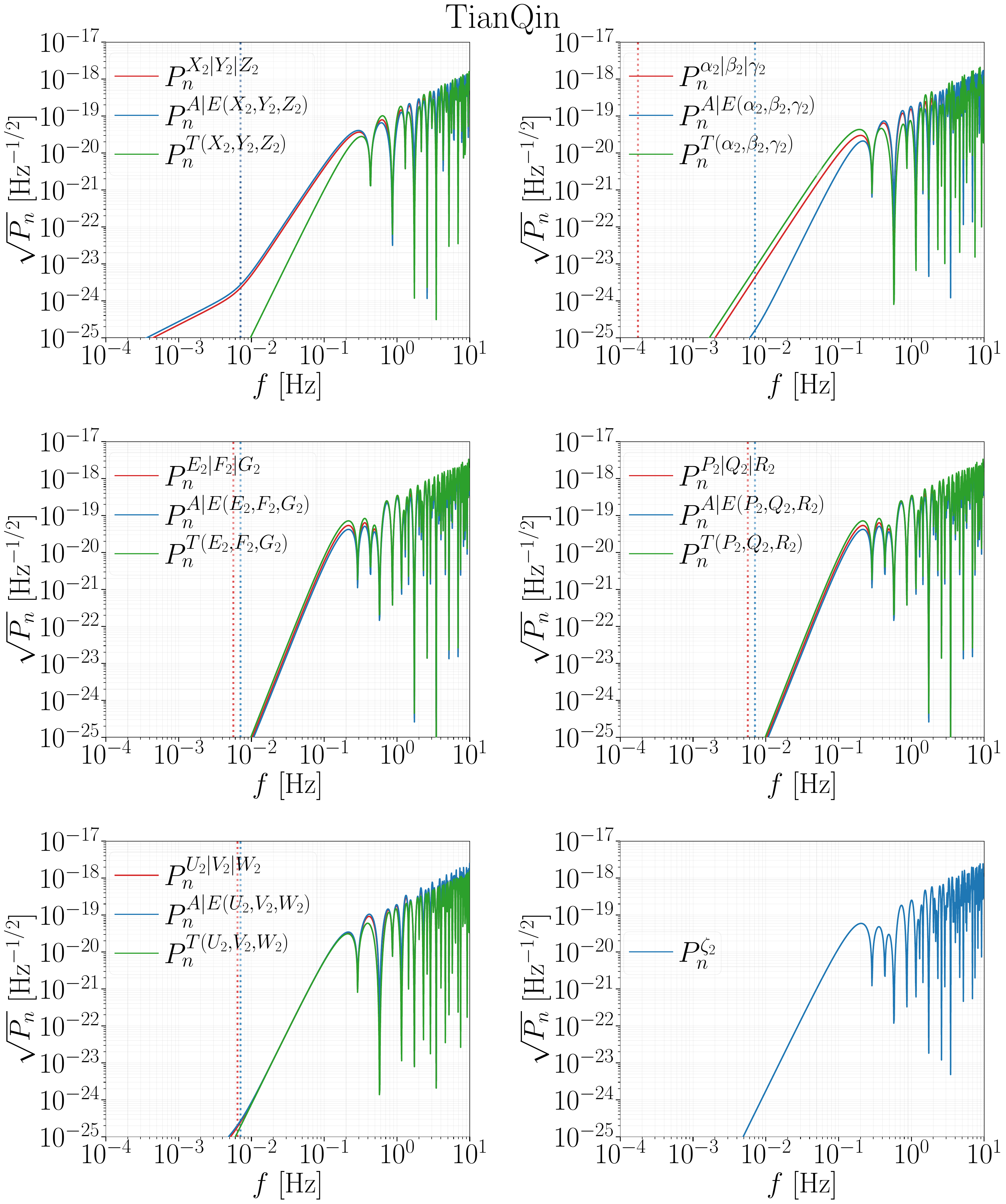}
\caption{Noise PSDs of the second‑generation TDI combinations for the TianQin mission, together with the PSDs of the corresponding orthogonal combinations.
Each solid line is accompanied by a vertical dotted line of the same color, marking  the noise-balance frequency $f_{n}$—the boundary between the proof-mass–dominated (low-frequency) and optical-path–dominated (high-frequency) noise regimes.}
\label{fig:PnTianQin}
\end{figure*}

The corresponding noise PSDs are
\begin{equation}
\label{eq:PnA_XYZ}
\begin{split}
&\frac{P_{n}^{A|E(X_{2},Y_{2},Z_{2})}}{16\sin^{2}u\sin^{2}2u}=\frac{P_{n}^{A|E(X_{1},Y_{1},Z_{1})}}{4\sin^{2}u}=P_{n}^{A|E(X_{0},Y_{0},Z_{0})}\\
=&4 \big(3+2 \cos u+\cos 2u\big)S_{y}^{\text{proof mass}}+ 2\big(2+\cos u\big)S_{y}^{\text{optical path}}\\
\sim& \begin{cases} 24 c^{\text{pm}}f^{-2}, & f\ll f_{n}^{A|E},\\
	2\big(2+\cos u\big)c^{\text{op}}f^{2},  &  f\gg f_{n}^{A|E},
\end{cases}
\end{split}
\end{equation}
where $f_{n}^{A|E}=(3.17\times10^{-3},4.35\times10^{-3},7.10\times10^{-3})$ Hz for LISA, Taiji, and TianQin, as shown in Figs. \ref{fig:PnLISA}, \ref{fig:PnTaiji} and \ref{fig:PnTianQin}.

\begin{figure}[htp]
\centering
\includegraphics[width=0.7\columnwidth]{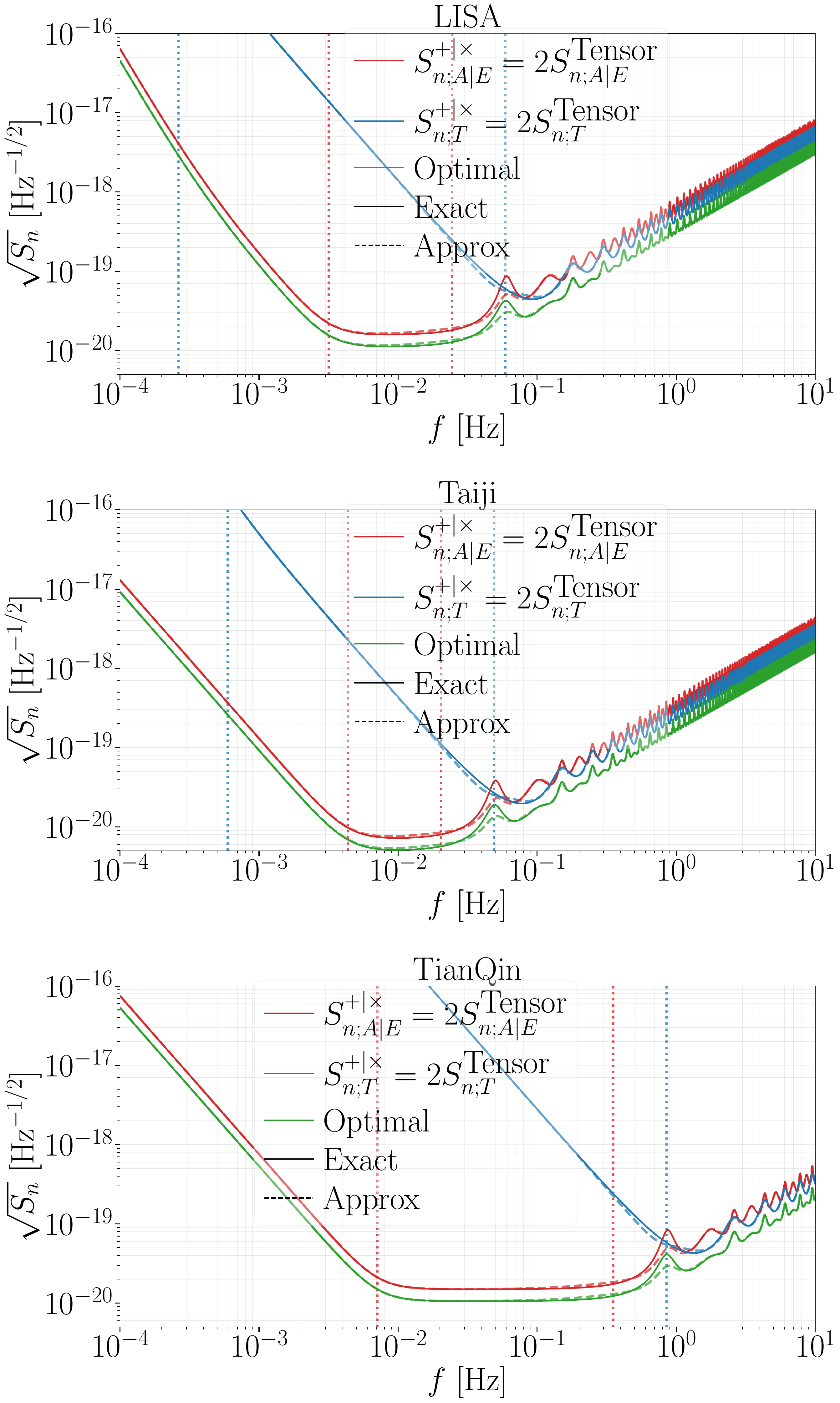}
\caption{Analytical (solid) and approximate (dashed) sensitivity curves for the orthogonal $A, E, T$ channels constructed from the first- and second-generation TDI Michelson, Sagnac-type ($\alpha, \beta, \gamma$), Monitor, Beacon, Relay, and fully symmetric Sagnac ($\zeta$) combinations.
Results are shown for the plus, cross, and composite tensor modes.
The composite tensor mode satisfies $R^{T}(f) = R^{+}(f) + R^{\times}(f) = 2R^{+|\times}(f)$.
The sensitivity curves of the orthogonal $A,E,T$ channels and the optimal sensitivity are independent of the TDI generation or the specific combination used.
Each solid curve for the orthogonal $A, E, T$ channels is accompanied by two vertical dotted lines of the same color: the left one marks the noise-balance frequency $f_{n}$—the boundary between the proof-mass–dominated (low-frequency) and optical-path–dominated (high-frequency) noise regimes;
the right one marks the response-transition frequency—the boundary between the response function’s low- and high-frequency behaviors.
}
\label{fig:SnAET}
\end{figure}

The corresponding sensitivity curves are
\begin{equation}
\label{eq:SnA_XYZ}
\begin{split}
&S^{+|\times}_{n;A|E(X_{2},Y_{2},Z_{2})}=S^{+|\times}_{n;A|E(X_{1},Y_{1},Z_{1})}=S^{+|\times}_{n;A|E(X_{0},Y_{0},Z_{0})}\\
\equiv&S^{+|\times}_{n;A|E(X,Y,Z)}\\
\sim& \begin{cases} \frac{20c^{\text{pm}}f^{-4}}{3\pi^{2}L^{2}}, & \hspace{-0.5em}f\ll f_{n}^{A|E},\\
	\frac{5(2+\cos u)c^{\text{op}}}{9\pi^{2}L^{2}}, & \hspace{-0.5em}f_{n}^{A|E}<f<\frac{1.28}{2\pi L},\\
	\frac{2(2+\cos u)c^{\text{op}}f^{2}}{1.06+0.43\cos u-0.35\cos 2u},  &  \hspace{-0.5em}f\gg \frac{1.28}{2\pi L}
\end{cases}
\end{split}
\end{equation}
Hence, the sensitivity curves of the orthogonal modes $A, E$ exhibit a flat section in the range $f_{n}^{A|E}<f<\frac{1.28}{2\pi L}$, as illustrated in Fig. \ref{fig:SnAET}.

The averaged response functions of orthogonal mode $T$, constructed from the Michelson combinations, are
\begin{equation}
\label{}
\begin{split}
&\frac{R^{+|\times}_{T(X_{2},Y_{2},Z_{2})}}{64\sin^{2}\frac{u}{2} \sin^{2}u\sin^{2}2u}=\frac{R^{+|\times}_{T(X_{1},Y_{1},Z_{1})}}{16\sin^{2}\frac{u}{2}\sin^{2}u}\\
=& \frac{R^{+|\times}_{T(X_{0},Y_{0},Z_{0})}}{4\sin^{2}\frac{u}{2}} \\
=& \frac{1}{12} + \ln 2 - \frac{5}{16u^{2}} + \text{Ci}(u) - \text{Ci}(2u) + \Big[-\frac{5}{8u^{3}} + \frac{1}{8u}\Big]\sin u \\
&+ \Big[\frac{3}{2}\text{Si}(u) - 3\text{Si}(2u) + \frac{3}{2}\text{Si}(3u)\Big]\sin u  + \Big[\frac{5}{16u^{3}} - \frac{1}{16u}\Big]\sin 2u \\
& + \Big[-\frac{1}{12} + \frac{3}{2}\ln\frac{4}{3} + \frac{5}{8u^{2}}\Big]\cos u + \Big[\frac{3}{2}\text{Ci}(u) - 3\text{Ci}(2u) + \frac{3}{2}\text{Ci}(3u)\Big]\cos u - \frac{5\cos 2u}{16u^{2}}\\
\sim& \begin{cases} \frac{1}{2016}u^{6}, & u\ll3.09,\\
	\frac{1}{12}+\ln 2 +\Big[-\frac{1}{12}+\frac{3}{2}\ln\frac{4}{3}\Big]\cos u,& u\gg3.09 \end{cases}\\
\approx& \frac{1}{2016}u^{6} \left(1+ \frac{u^{6}}{1565+702 \cos u}\right)^{-1},
\end{split}
\end{equation}
as shown in Fig. \ref{fig:Rpc}.

The corresponding noise PSDs are
\begin{equation}
\label{eq:PnT_XYZ}
\begin{split}
&\frac{P_{n}^{T(X_{2},Y_{2},Z_{2})}}{64\sin^{2}\frac{u}{2} \sin^{2}u\sin^{2}2u}=\frac{P_{n}^{T(X_{1},Y_{1},Z_{1})}}{16\sin^{2}\frac{u}{2}\sin^{2}u}\\
=& \frac{P_{n}^{T(X_{0},Y_{0},Z_{0})}}{4\sin^{2}\frac{u}{2}} \\
=& 8 \sin ^{2}\left( \frac{u}{2}\right) S_{y}^{\text{proof mass}}+ 2S_{y}^{\text{optical path}}\\
\sim& \begin{cases} 8\pi^{2}L^{2} c^{\text{pm}}, & f\ll f_{n}^{T},\\
	2c^{\text{op}}f^{2},  &  f\gg f_{n}^{T},
\end{cases}		
\end{split}
\end{equation}
where $f_{n}^{T}=(2.63\times10^{-4},5.94\times10^{-4},9.15\times10^{-5})$ Hz for LISA, Taiji, and TianQin, respectively, as shown in Figs. \ref{fig:PnLISA}, \ref{fig:PnTaiji} and \ref{fig:PnTianQin}.

The corresponding sensitivity curves are
\begin{equation}
\label{eq:SnT_XYZ}
\begin{split}
&S^{+|\times}_{n;T(X_{2},Y_{2},Z_{2})}=S^{+|\times}_{n;T(X_{1},Y_{1},Z_{1})}=S^{+|\times}_{n;T(X_{0},Y_{0},Z_{0})}\\
\equiv&S^{+|\times}_{n;T(X,Y,Z)}\\
\sim& \begin{cases} \frac{252}{\pi^{4}L^{4}}c^{\text{pm}}f^{-6}, & f\ll f_{n}^{T},\\
	\frac{63}{\pi^{6}L^{6}}c^{\text{op}}f^{-4}, & f_{n}^{T}<f<\frac{3.09}{2\pi L},\\
	\frac{2c^{\text{op}}}{0.78+0.35 \cos u}f^{2},  &  f\gg \frac{3.09}{2\pi L}.
\end{cases}
\end{split}
\end{equation}
Hence, unlike the $A$ and $E$ channels, the sensitivity curve of the orthogonal mode $T$ does not exhibit a flat section; instead, it degrades as $f^{-4}$ and then $f^{-6}$ when $f<\frac{3.09}{2\pi L}$, as illustrated in Fig. \ref{fig:SnAET}.

\subsection{Sagnac-type combinations ($\alpha,\beta,\gamma$)}

The first-generation $\alpha$ combination is
\begin{equation}
\label{eq:alpha1}
\begin{split}
\alpha_{1} =& \mathcal{P}_{1321} - \mathcal{P}_{1231}\\
=&[y_{13}+\mathcal{D}_{13}y_{32}+\mathcal{D}_{132}y_{21}]\\
&-[y_{12}+\mathcal{D}_{12}y_{23}+\mathcal{D}_{123}y_{31}].
\end{split}
\end{equation}
The second-generation $\alpha$ combination is
\begin{equation}
\label{eq:alpha2}
\begin{split}
\alpha_{2} =& \mathcal{P}_{1321231231321} - \mathcal{P}_{1231321321231}\\
=& y_{13}+\mathcal{D}_{13} y_{32} + \mathcal{D}_{132} y_{21} + \mathcal{D}_{1321} y_{12} + \mathcal{D}_{13212} y_{23}  + \mathcal{D}_{132123} y_{31} + \mathcal{D}_{1321231} y_{12}  \\
& + \mathcal{D}_{13212312} y_{23} + \mathcal{D}_{132123123} y_{31} + \mathcal{D}_{1321231231} y_{13}  + \mathcal{D}_{13212312313} y_{32} + \mathcal{D}_{132123123132} y_{21} \\
& - \Big[y_{12}+\mathcal{D}_{12} y_{23} + \mathcal{D}_{123} y_{31} + \mathcal{D}_{1231} y_{13}  + \mathcal{D}_{12313} y_{32} + \mathcal{D}_{123132} y_{21} + \mathcal{D}_{1231321} y_{13}  \\
& + \mathcal{D}_{12313213} y_{32} + \mathcal{D}_{123132132} y_{21} + \mathcal{D}_{1231321321} y_{12}  + \mathcal{D}_{12313213212} y_{23}+ \mathcal{D}_{123132132123} y_{31}\Big]. 
\end{split}
\end{equation}
The light paths of $\alpha_{2}$ are illustrated in Fig. \ref{fig:path}.

\subsubsection{Sensitivity curves of $\alpha,\beta,\gamma$}

From Eqs. \eqref{eq:Ra} and \eqref{eq:alpha2},
the sky- and polarization-averaged response functions of $\alpha$ combinations are
\begin{equation}
\label{eq:Ralpha}
\begin{split}
&\frac{R^{+|\times}_{\alpha_{2}|\beta_{2}|\gamma_{2}}}{16\sin^{2} \frac{3u}{2}\sin^{2}3u}=R^{+|\times}_{\alpha_{1}|\beta_{1}|\gamma_{1}} \\
=& \frac{7}{12} + 3\ln\frac{4}{3} + \ln 2 - \frac{27}{16u^{2}} + 4\text{Ci}(u) - 7\text{Ci}(2u) + 3\text{Ci}(3u) - \Big[\frac{9}{4u^{3}} +\frac{17}{4u}\Big]\sin u \\
&+ \Big[\frac{27}{16u^{3}} + \frac{49}{16u}\Big]\sin 2u -\Big[\frac{3}{8u^{3}} + \frac{5}{8u}\Big]\sin 3u  + \Big[\frac{3}{2}\text{Si}(u) - 3\text{Si}(2u) + \frac{3}{2}\text{Si}(3u)\Big]\sin 3u \\  
& + \Big[-\frac{1}{2} + 2\ln 2 + \frac{3}{u^{2}}\Big]\cos u + \Big[2\text{Ci}(u) - 2\text{Ci}(2u)\Big]\cos u - \frac{27\cos 2u}{16u^{2}} \\  
& + \Big[-\frac{1}{12} + \frac{3}{2}\ln\frac{4}{3} + \frac{3}{8u^{2}}\Big]\cos 3u + \Big[\frac{3}{2}\text{Ci}(u) - 3\text{Ci}(2u) + \frac{3}{2}\text{Ci}(3u)\Big]\cos 3u\\
\sim& \begin{cases} \frac{3}{5}u^{4}, & u\ll1.37,\\
	\frac{7}{12}+\ln 2+3\ln\frac{4}{3} +\Big[-\frac{1}{2}+2\ln 2\Big]\cos u +\Big[-\frac{1}{12}+\frac{3}{2}\ln\frac{4}{3}\Big]\cos 3u,
	& u\gg1.37 \end{cases}\\
\approx& \frac{3}{5}u^{4} \left(1+ \frac{u^{4}}{3.57+ 1.48\cos u+0.58 \cos3u}\right)^{-1},
\end{split}
\end{equation}
as shown in Fig. \ref{fig:Rpc}.

The noise terms of the second-generation $\alpha$ combination are analyzed in the Appendix \ref{app:noise}.
The corresponding noise PSDs are
\begin{equation}
\label{eq:Pn_alphabetagamma}
\begin{split}
&\frac{P_{n}^{\alpha_{2}|\beta_{2}|\gamma_{2}}}{16\sin^{2} \frac{3u}{2}\sin^{2}3u}=P_{n}^{\alpha_{1}|\beta_{1}|\gamma_{1}} \\
=& 8 \big(5 + 4\cos u + 2\cos2u\big)\sin^{2}\left(\frac{u}{2}\right)  S_{y}^{\text{proof mass}} + 6 S_{y}^{\text{optical path}}\\
\sim& \begin{cases} 88\pi^{2}L^{2} c^{\text{pm}}, & f\ll f_{n}^{\alpha|\beta|\gamma},\\
	6c^{\text{op}}f^{2},  &  f\gg f_{n}^{\alpha|\beta|\gamma},
\end{cases}
\end{split}
\end{equation}
where $f_{n}^{\alpha|\beta|\gamma}=(5.04\times10^{-4},1.14\times10^{-3},1.75\times10^{-4})$ Hz for LISA, Taiji, and TianQin, respectively, as shown in Figs. \ref{fig:PnLISA}, \ref{fig:PnTaiji} and \ref{fig:PnTianQin}.

The corresponding sensitivity curves are
\begin{equation}
\label{eq:Sn_alphabetagamma}
\begin{split}
&S^{+|\times}_{n;\alpha_{2}|\beta_{2}|\gamma_{2}}=S^{+|\times}_{n;\alpha_{1}|\beta_{1}|\gamma_{1}}\\
\equiv& S^{+|\times}_{n;\alpha|\beta|\gamma}\\
\sim& \begin{cases} \frac{55}{6\pi^{2}L^{2}}c^{\text{pm}}f^{-4}, & \hspace{-2em}f\ll f_{n}^{\alpha|\beta|\gamma},\\
	\frac{5c^{\text{op}}}{8\pi^{4}L^{4}}f^{-2}, &\hspace{-2em} f_{n}^{\alpha|\beta|\gamma}<f<\frac{1.37}{2\pi L},\\
	\frac{6c^{\text{op}}}{2.14+0.89\cos u+0.35 \cos 3u}f^{2},  &  f\gg \frac{1.37}{2\pi L}.
\end{cases}
\end{split}
\end{equation}
Hence, unlike other combinations, the sensitivity curve of the $(\alpha,\beta,\gamma)$ does not exhibit a flat section; 
instead, it degrades as $f^{-2}$ and then $f^{-4}$ when $f<\frac{1.37}{2\pi L}$, as illustrated in Fig. \ref{fig:Sn}.

\subsubsection{Sensitivity curves of $A, E, T$}

From Eqs. \eqref{eq:AET} and \eqref{eq:alpha2}, the sky- and polarization-averaged response functions and noise PSDs of the orthogonal modes $A$, $E$, $T$, constructed from the $\alpha,\beta,\gamma$ combinations, are
\begin{equation}
\label{eq:RPnAE_alphabetagamma}
\begin{split}
\frac{R^{+|\times}_{A|E(\alpha_{2},\beta_{2},\gamma_{2})}}{64\sin^{2} \frac{u}{2} \sin^{2} \frac{3u}{2}\sin^{2}3u}=& \frac{R^{+|\times}_{A|E(\alpha_{1},\beta_{1},\gamma_{1})}}{4\sin^{2} \frac{u}{2}}
= R^{+|\times}_{A|E(X_{0},Y_{0},Z_{0})},\\
\frac{P_{n}^{A|E(\alpha_{2},\beta_{2},\gamma_{2})}}{64\sin^{2} \frac{u}{2} \sin^{2} \frac{3u}{2}\sin^{2}3u}=& \frac{P_{n}^{A|E(\alpha_{1},\beta_{1},\gamma_{1})}}{4\sin^{2} \frac{u}{2}}
= P_{n}^{A|E(X_{0},Y_{0},Z_{0})},
\end{split}
\end{equation}
and
\begin{equation}
\label{eq:RPnT_alphabetagamma}
\begin{split}
\frac{R^{+|\times}_{T(\alpha_{2},\beta_{2},\gamma_{2})}}{16(1+2\cos u)^{2}\sin^{2}\frac{3u}{2} \sin^{2}3u}=& \frac{R^{+|\times}_{T(\alpha_{1},\beta_{1},\gamma_{1})}}{(1+2\cos u)^{2}}
= \frac{R^{+|\times}_{T(X_{0},Y_{0},Z_{0})}}{4\sin^{2}\frac{u}{2}},\\
\frac{P_{n}^{T(\alpha_{2},\beta_{2},\gamma_{2})}}{16(1+2\cos u)^{2}\sin^{2}\frac{3u}{2} \sin^{2}3u}=& \frac{P_{n}^{T(\alpha_{1},\beta_{1},\gamma_{1})}}{(1+2\cos u)^{2}}
= \frac{P_{n}^{T(X_{0},Y_{0},Z_{0})}}{4\sin^{2}\frac{u}{2}},
\end{split}
\end{equation}
as shown in Figs. \ref{fig:Rpc}, \ref{fig:PnLISA}, \ref{fig:PnTaiji} and \ref{fig:PnTianQin}.

Thus, the sensitivity curves of the orthogonal modes $A$, $E$, $T$, constructed from the $\alpha,\beta,\gamma$ combinations, are equal to those obtained from Michelson combinations, 
\begin{equation}
\label{eq:SnAET_alphabetagamma}
\begin{split}
S^{+|\times}_{n;A|E(\alpha,\beta,\gamma)}=& S^{+|\times}_{n;A|E(X,Y,Z)},\\
S^{+|\times}_{n;T(\alpha,\beta,\gamma)}=& S^{+|\times}_{n;T(X,Y,Z)},
\end{split}
\end{equation}
as shown in Fig. \ref{fig:SnAET}.

\subsection{Monitor combinations (E, F, G)}

The first-generation $E$ combination is
\begin{equation}
\label{eq:E1}
\begin{split}
E_{1} =& [\mathcal{D}_{323}y_{12}+ \mathcal{P}_{1323}] - [\mathcal{D}_{232}y_{13}+ \mathcal{P}_{1232}]\\
=& \left[ \mathcal{D}_{323} y_{12}+ y_{13} + \mathcal{D}_{13} y_{32} + \mathcal{D}_{132} y_{23}\right]  \\
&- \left[ \mathcal{D}_{232} y_{13}+ y_{12} + \mathcal{D}_{12} y_{23} + \mathcal{D}_{123} y_{32} \right].
\end{split}
\end{equation}
The second-generation $E$ combination is
\begin{equation}
\label{eq:E2}
\begin{split}
E_{2} =& -(\mathcal{D}_{1231321}-1)(\mathcal{D}_{1321}-1)\mathcal{D}_{23}\mathcal{D}_{23}X_{2} \\
&+(1-\mathcal{D}_{12131})\Big[(1+ \mathcal{D}_{2321})\alpha_{2}- \mathcal{D}_{231}\beta_{2}- \mathcal{D}_{23}\mathcal{D}_{12} \gamma_{2}\Big].
\end{split}
\end{equation}

\subsubsection{Sensitivity curves of E, F, G}

From Eqs. \eqref{eq:Ra} and \eqref{eq:E2},
the sky- and polarization-averaged response functions of Monitor combinations are
\begin{equation}
\label{eq:REFG}
\begin{split}
&\frac{R^{+|\times}_{E_{2}|F_{2}|G_{2}}}{256\sin^{2} \frac{u}{2} \sin^{2} \frac{3u}{2}\sin^{2}2u\sin^{2}3u} =  \frac{R^{+|\times}_{E_{1}|F_{1}|G_{1}}}{4\sin^{2} \frac{u}{2}} \\
=& \frac{5}{12} + \frac{3}{2}\ln\frac{4}{3} + 2\ln 2 - \frac{11}{16u^{2}} + \frac{7}{2}\text{Ci}(u) - 5\text{Ci}(2u) + \frac{3}{2}\text{Ci}(3u) + \Big[-\frac{25}{16u^{3}} -\frac{27}{16u}\Big]\sin u \\
&+ \Big[\frac{3}{2}\text{Si}(u) - 3\text{Si}(2u) + \frac{3}{2}\text{Si}(3u)\Big]\sin u  + \Big[\frac{11}{16u^{3}} + \frac{9}{16u}\Big]\sin 2u + \Big[\frac{5}{16u^{3}} - \frac{1}{16u}\Big]\sin 3u \\
& + \Big[\frac{1}{12} + \frac{3}{2}\ln\frac{4}{3} + 2\ln 2 + \frac{15}{16u^{2}}\Big]\cos u + \Big[\frac{7}{2}\text{Ci}(u) - 5\text{Ci}(2u) + \frac{3}{2}\text{Ci}(3u)\Big]\cos u \\
& - \frac{11\cos 2u}{16u^{2}} - \frac{5\cos 3u}{16u^{2}}\\
\sim& \begin{cases} \frac{3}{5}u^{2}, & u\ll1.76,\\
	\frac{5}{12}+2\ln 2+\frac{3}{2}\ln\frac{4}{3} +\Big[\frac{1}{12}+2\ln 2+\frac{3}{2}\ln\frac{4}{3}\Big]\cos u,
	& u\gg1.76 \end{cases}\\
\approx& \frac{3}{5}u^{2} \left(1+ \frac{u^{2}}{3.72+ 3.17 \cos u}\right)^{-1},
\end{split}
\end{equation}
as shown in Fig. \ref{fig:Rpc}.

The corresponding noise PSDs are
\begin{equation}
\label{eq:PnE2}
\begin{split}
&\frac{P_{n}^{E_{2}|F_{2}|G_{2}}}{256\sin^{2} \frac{u}{2} \sin^{2} \frac{3u}{2}\sin^{2}2u\sin^{2}3u} =  \frac{P_{n}^{E_{1}|F_{1}|G_{1}}}{4\sin^{2} \frac{u}{2}}\\
=& 4\big(3+\cos u\big) S_{y}^{\text{proof mass}} + 2\big(3+2\cos u\big) S_{y}^{\text{optical path}}\\
\sim& \begin{cases} 16 c^{\text{pm}} f^{-2}, & f\ll f_{n}^{E|F|G},\\
	2\big(3+2\cos u\big)c^{\text{op}}f^{2},  & f\gg f_{n}^{E|F|G},
\end{cases}
\end{split}
\end{equation}
where $f_{n}^{E|F|G}=(2.52\times10^{-3},3.46\times10^{-3},5.65\times10^{-3})$ Hz for LISA, Taiji, and TianQin, respectively, as shown in Figs. \ref{fig:PnLISA}, \ref{fig:PnTaiji} and \ref{fig:PnTianQin}.

The corresponding sensitivity curves are
\begin{equation}
\label{eq:Sn_EFG}
\begin{split}
&S^{+|\times}_{n;E_{2}|F_{2}|G_{2}}=S^{+|\times}_{n;E_{1}|F_{1}|G_{1}}\equiv S^{+|\times}_{n;E|F|G}\\
\sim& \begin{cases} \frac{20}{3\pi^{2}L^{2}}c^{\text{pm}}f^{-4}, & f\ll f_{n}^{E|F|G},\\
	\frac{5(3+2\cos u)c^{\text{op}}}{6\pi^{2}L^{2}}, & f_{n}^{E|F|G}<f<\frac{1.76}{2\pi L},\\
	\frac{2(3+2\cos u)c^{\text{op}}}{2.23+1.9\cos u}f^{2},  &  f\gg \frac{1.76}{2\pi L}.
\end{cases}
\end{split}
\end{equation}
Hence, the sensitivity curves of the Monitor combinations exhibit a flat section in the range $f_{n}^{E|F|G}<f<\frac{1.76}{2\pi L}$, as illustrated in Fig. \ref{fig:Sn}.

\subsubsection{Sensitivity curves of $A, E, T$}

From Eqs. \eqref{eq:AET} and \eqref{eq:E2}, the sky- and polarization-averaged response functions and noise PSDs of the orthogonal modes $A$, $E$, $T$, constructed from the Monitor combinations, are
\begin{equation}
\label{eq:RPnAE_EFG}
\begin{split}
&\frac{R^{+|\times}_{A|E(E_{2},F_{2},G_{2})}}{256\sin^{2} \frac{u}{2} \sin^{2} \frac{3u}{2}\sin^{2}2u\sin^{2}3u}
= \frac{R^{+|\times}_{A|E(E_{1},F_{1},G_{1})}}{4\sin^{2} \frac{u}{2}}
= R^{+|\times}_{A|E(X_{0},Y_{0},Z_{0})},\\
&\frac{P_{n}^{A|E(E_{2},F_{2},G_{2})}}{256\sin^{2} \frac{u}{2} \sin^{2} \frac{3u}{2}\sin^{2}2u\sin^{2}3u}
= \frac{P_{n}^{A|E(E_{1},F_{1},G_{1})}}{4\sin^{2} \frac{u}{2}}
= P_{n}^{A|E(X_{0},Y_{0},Z_{0})}.
\end{split}
\end{equation}
and
\begin{equation}
\label{eq:RPnT_EFG}
\begin{split}
&\frac{R^{+|\times}_{T(E_{2},F_{2},G_{2})}}{256(5+4\cos u)\sin^{2} \frac{u}{2} \sin^{2} \frac{3u}{2}\sin^{2}2u\sin^{2}3u}
= \frac{R^{+|\times}_{T(E_{1},F_{1},G_{1})}}{4(5+4\cos u)\sin^{2}\frac{u}{2}}
= \frac{R^{+|\times}_{T(X_{0},Y_{0},Z_{0})}}{4\sin^{2}\frac{u}{2}},\\
&\frac{P_{n}^{T(E_{2},F_{2},G_{2})}}{256(5+4\cos u)\sin^{2} \frac{u}{2} \sin^{2} \frac{3u}{2}\sin^{2}2u\sin^{2}3u}
= \frac{P_{n}^{T(E_{1},F_{1},G_{1})}}{4(5+4\cos u)\sin^{2}\frac{u}{2}}
= \frac{P_{n}^{T(X_{0},Y_{0},Z_{0})}}{4\sin^{2}\frac{u}{2}},
\end{split}
\end{equation}
as shown in Figs. \ref{fig:Rpc}, \ref{fig:PnLISA}, \ref{fig:PnTaiji} and \ref{fig:PnTianQin}.

Thus, the sensitivity curves of the orthogonal modes $A$, $E$, $T$, constructed from the Monitor combinations, are equal to those obtained from Michelson combinations, 
\begin{equation}
\label{eq:SnAET_EFG}
\begin{split}
S^{+|\times}_{n;A|E(E,F,G)}=& S^{+|\times}_{n;A|E(X,Y,Z)},\\
S^{+|\times}_{n;T(E,F,G)}=& S^{+|\times}_{n;T(X,Y,Z)},
\end{split}
\end{equation}
as shown in Fig. \ref{fig:SnAET}.

\subsection{Beacon combinations (P, Q, R)}

The first-generation $P$ combination is
\begin{equation}
\label{eq:P1}
\begin{split}
P_{1} =& [\mathcal{D}_{31}y_{21}+\mathcal{D}_{21}\mathcal{P}_{3231}] - [\mathcal{D}_{21}y_{31}+\mathcal{D}_{31}\mathcal{P}_{2321}]\\
=& \left[\mathcal{D}_{31} y_{21} + \mathcal{D}_{21} (y_{32} + \mathcal{D}_{32}y_{23} + \mathcal{D}_{323}y_{31})\right] \\
& - \left[ \mathcal{D}_{21} y_{31} + \mathcal{D}_{31} (y_{23} + \mathcal{D}_{23} y_{32} + \mathcal{D}_{232} y_{21} )\right] .
\end{split}
\end{equation}
The second-generation $P$ combination is
\begin{equation}
\label{eq:P2}
\begin{split}
P_{2} =& (\mathcal{D}_{1231321}-1)(\mathcal{D}_{1321}-1)\mathcal{D}_{23}X_{2}\\
&+ (1-\mathcal{D}_{12131})\Big[-(\mathcal{D}_{23}+ \mathcal{D}_{321})\alpha_{2}+ \mathcal{D}_{31}\beta_{2}+ \mathcal{D}_{12} \gamma_{2}\Big].
\end{split}
\end{equation}

\subsubsection{Sensitivity curves of P, Q, R}

From Eqs. \eqref{eq:Ra} and \eqref{eq:P2},
the sky- and polarization-averaged response functions, noise PSDs, and sensitivity curves of the Beacon combinations are equal to those of the Monitor combinations:
\begin{equation}
\label{}
\begin{split}
R^{+|\times}_{P_{2}|Q_{2}|R_{2}} = R^{+|\times}_{E_{2}|F_{2}|G_{2}},  \quad
R^{+|\times}_{P_{1}|Q_{1}|R_{1}} =   R^{+|\times}_{E_{1}|F_{1}|G_{1}},
\end{split}
\end{equation}
\begin{equation}
\label{eq:PnP2}
\begin{split}
P_{n}^{P_{2}|Q_{2}|R_{2}} = P_{n}^{E_{2}|F_{2}|G_{2}}, \quad
P_{n}^{P_{1}|Q_{1}|R_{1}} = P_{n}^{E_{1}|F_{1}|G_{1}},
\end{split}
\end{equation}
and
\begin{equation}
\label{}
\begin{split}
S^{+|\times}_{n;P|Q|R} =& S^{+|\times}_{n;E|F|G}.
\end{split}
\end{equation}

\subsubsection{Sensitivity curves of $A, E, T$}

From Eqs. \eqref{eq:AET} and \eqref{eq:P2}, the sky- and polarization-averaged response functions and noise PSDs of the orthogonal modes $A$, $E$, $T$, constructed from the Beacon combinations, are equal to those obtained from the Monitor combinations:
\begin{equation}
\label{eq:RPnAE_PQR}
\begin{split}
R^{+|\times}_{A|E(P_{2},Q_{2},R_{2})} =& R^{+|\times}_{A|E(E_{2},F_{2},G_{2})},\quad  R^{+|\times}_{A|E(P_{1},Q_{1},R_{1})} = R^{+|\times}_{A|E(E_{1},F_{1},G_{1})},\\
P_{n}^{A|E(P_{2},Q_{2},R_{2})} =& P_{n}^{A|E(E_{2},F_{2},G_{2})}, \quad 
P_{n}^{A|E(P_{1},Q_{1},R_{1})} = P_{n}^{A|E(E_{1},F_{1},G_{1})},
\end{split}
\end{equation}
and
\begin{equation}
\label{eq:RPnT_PQR}
\begin{split}
R^{+|\times}_{T(P_{2},Q_{2},R_{2})} =& R^{+|\times}_{T(E_{2},F_{2},G_{2})}, \quad R^{+|\times}_{T(P_{1},Q_{1},R_{1})} = R^{+|\times}_{T(E_{1},F_{1},G_{1})}, \\
P_{n}^{T(P_{2},Q_{2},R_{2})} =& P_{n}^{T(E_{2},F_{2},G_{2})}, \quad
P_{n}^{T(P_{1},Q_{1},R_{1})} = P_{n}^{T(E_{1},F_{1},G_{1})}.
\end{split}
\end{equation}
Thus, the sensitivity curves of the orthogonal modes $A$, $E$, $T$, constructed from the Beacon combinations, are equal to those obtained from the Monitor combinations or the Michelson combinations:
\begin{equation}
\label{eq:SnAET_PQR}
\begin{split}
S^{+|\times}_{n;A|E(P,Q,R)} =& S^{+|\times}_{n;A|E(E,F,G)}=S^{+|\times}_{n;A|E(X,Y,Z)},\\
S^{+|\times}_{n;T(P,Q,R)}=& S^{+|\times}_{n;T(E,F,G)}= S^{+|\times}_{n;T(X,Y,Z)}.
\end{split}
\end{equation}

\subsection{Relay combinations (U, V, W)}

The first-generation $U$ combination is
\begin{equation}
\label{eq:U1}
\begin{split}
U_{1} =& \mathcal{P}_{23213} - \mathcal{P}_{21323}\\
=& y_{23} + \mathcal{D}_{23} y_{32} + \mathcal{D}_{232} y_{21} + \mathcal{D}_{2321} y_{13} \\
&- \left[ y_{21} + \mathcal{D}_{21} y_{13} + \mathcal{D}_{213} y_{32} + \mathcal{D}_{2132} y_{23} \right].
\end{split}
\end{equation}
The second-generation $U$ combination is
\begin{equation}
\label{eq:U2}
\begin{split}
U_{2} =& -\beta_{2}+ \mathcal{D}_{23} \gamma_{2}.
\end{split}
\end{equation}

\subsubsection{Sensitivity curves of U, V, W}

From Eqs. \eqref{eq:Ra} and \eqref{eq:U2},
the sky- and polarization-averaged response functions of Relay combinations are
\begin{equation}
\label{}
\begin{split}
&\frac{R^{+|\times}_{U_{2}|V_{2}|W_{2}}}{64\sin^{2} \frac{u}{2} \sin^{2} \frac{3u}{2}\sin^{2}3u} =  \frac{R^{+|\times}_{U_{1}|V_{1}|W_{1}}}{4\sin^{2} \frac{u}{2}}\\
=& \frac{3}{4} + \frac{3}{2}\ln\frac{4}{3} + 2\ln 2 - \frac{29}{32u^{2}} + \frac{7}{2}\text{Ci}(u) - 5\text{Ci}(2u) + \frac{3}{2}\text{Ci}(3u)- \Big[\frac{11}{8u^{3}} + \frac{27}{8u}\Big]\sin u \\
&+ \Big[-\frac{3}{2}\text{Si}(u) + 3\text{Si}(2u) - \frac{3}{2}\text{Si}(3u)\Big]\sin u  + \Big[\frac{3}{4u^{3}} + \frac{1}{4u}\Big]\sin 2u \\
&+ \Big[-\frac{3}{2}\text{Si}(u) + 3\text{Si}(2u) - \frac{3}{2}\text{Si}(3u)\Big]\sin 2u  + \Big[\frac{1}{2u^{3}} + \frac{1}{4u}\Big]\sin 3u + \Big[\frac{5}{32u^{3}} - \frac{1}{32u}\Big]\sin 4u\\
&+ \Big[\frac{7}{12} + 3\ln 2 + \frac{3}{8u^{2}}\Big]\cos u   + \Big[3\text{Ci}(u) - 3\text{Ci}(2u)\Big]\cos u \\
& + \Big[\frac{1}{6} - \frac{3}{2}\ln\frac{4}{3} + \ln 2 - \frac{17}{16u^{2}}\Big]\cos 2u + \Big[-\frac{1}{2}\text{Ci}(u) + 2\text{Ci}(2u) - \frac{3}{2}\text{Ci}(3u)\Big]\cos 2u \\
& - \frac{\cos 3u}{2u^{2}} - \frac{5\cos 4u}{32u^{2}} \\
\sim& \begin{cases} \frac{9}{5}u^{2}, & u\ll1.28,\\
	\frac{3}{4}+2\ln 2+\frac{3}{2}\ln\frac{4}{3} +\Big[\frac{7}{12}+3\ln 2\Big]\cos u +\Big[\frac{1}{6}-\frac{3}{2}\ln\frac{4}{3}+\ln 2\Big]\cos 2u,
	& u\gg1.28 \end{cases}\\
\approx& \frac{9}{5}u^{2} \left(1+ \frac{u^{2}}{1.43+ 1.48 \cos u+ 0.24 \cos 2u}\right)^{-1},
\end{split}
\end{equation}
as shown in Fig. \ref{fig:Rpc}.

The corresponding noise PSDs are
\begin{equation}
\label{eq:PnU2}
\begin{split}
&\frac{P_{n}^{U_{2}|V_{2}|W_{2}}}{64\sin^{2} \frac{u}{2} \sin^{2} \frac{3u}{2}\sin^{2}3u} =  \frac{P_{n}^{U_{1}|V_{1}|W_{1}}}{4\sin^{2} \frac{u}{2}}\\
=& 4\big(5+5\cos u + 2\cos2u\big) S_{y}^{\text{proof mass}} + 2\big(4+4\cos u + \cos2u\big) S_{y}^{\text{optical path}}\\
\sim& \begin{cases} 48 c^{\text{pm}} f^{-2}, & f\ll f_{n}^{U|V|W},\\
	2\big(4+4\cos u + \cos2u\big)c^{\text{op}}f^{2},  &  f\gg f_{n}^{U|V|W},
\end{cases}
\end{split}
\end{equation}
where $f_{n}^{U|V|W}=(2.87\times10^{-3},3.93\times10^{-3},6.42\times10^{-3})$ Hz for LISA, Taiji, and TianQin, respectively, as shown in Figs. \ref{fig:PnLISA}, \ref{fig:PnTaiji} and \ref{fig:PnTianQin}.

The corresponding sensitivity curves are
\begin{equation}
\label{eq:Sn_UVW}
\begin{split}
&S^{+|\times}_{n;U_{2}|V_{2}|W_{2}}=S^{+|\times}_{n;U_{1}|V_{1}|W_{1}}\\
\equiv&S^{+|\times}_{n;U|V|W}\\
\sim& \begin{cases} \frac{20}{3\pi^{2}L^{2}}c^{\text{pm}}f^{-4}, & \hspace{-1.5em}f\ll f_{n}^{U|V|W},\\
	\frac{5(4+4\cos u+\cos 2u)c^{\text{op}}}{18\pi^{2}L^{2}}, & \hspace{-1.5em}f_{n}^{U|V|W}<f<\frac{1.28}{2\pi L},\\
	\frac{2(4+4\cos u+\cos 2u)c^{\text{op}}}{2.57+2.66\cos u+0.43\cos2u}f^{2},  & f\gg \frac{1.28}{2\pi L}.
\end{cases}
\end{split}
\end{equation}
Hence, the sensitivity curves of the Relay combinations exhibit a flat section in the range $ f_{n}^{U|V|W}<f<\frac{1.28}{2\pi L}$, as illustrated in Fig. \ref{fig:Sn}.

\subsubsection{Sensitivity curves of $A, E, T$}

From Eqs. \eqref{eq:AET} and \eqref{eq:U2}, the sky- and polarization-averaged response functions and noise PSDs of the orthogonal modes $A$, $E$, $T$, constructed from the Relay combinations, are
\begin{equation}
\label{eq:RPnAE_UVW}
\begin{split}
&\frac{R^{+|\times}_{A|E(U_{2},V_{2},W_{2})}}{64(2+\cos u)\sin^{2} \frac{u}{2} \sin^{2} \frac{3u}{2}\sin^{2}3u}
= \frac{R^{+|\times}_{A|E(U_{1},V_{1},W_{1})}}{4(2+\cos u)\sin^{2} \frac{u}{2}}
= R^{+|\times}_{A|E(X_{0},Y_{0},Z_{0})},\\
&\frac{P_{n}^{A|E(U_{2},V_{2},W_{2})}}{64(2+\cos u)\sin^{2} \frac{u}{2} \sin^{2} \frac{3u}{2}\sin^{2}3u}
= \frac{P_{n}^{A|E(U_{1},V_{1},W_{1})}}{4(2+\cos u)\sin^{2} \frac{u}{2}}
= P_{n}^{A|E(X_{0},Y_{0},Z_{0})},
\end{split}
\end{equation}
and
\begin{equation}
\label{eq:RPnT_UVW}
\begin{split}
&\frac{R^{+|\times}_{T(U_{2},V_{2},W_{2})}}{64\sin^{4}\frac{3u}{2} \sin^{2}3u}
=\frac{R^{+|\times}_{T(U_{1},V_{1},W_{1})}}{4\sin^{2}\frac{3u}{2}}
= \frac{R^{+|\times}_{T(X_{0},Y_{0},Z_{0})}}{4\sin^{2}\frac{u}{2}},\\
&\frac{P_{n}^{T(U_{2},V_{2},W_{2})}}{64\sin^{4}\frac{3u}{2} \sin^{2}3u}
= \frac{P_{n}^{T(U_{1},V_{1},W_{1})}}{4\sin^{2}\frac{3u}{2}}
= \frac{P_{n}^{T(X_{0},Y_{0},Z_{0})}}{4\sin^{2}\frac{u}{2}}.
\end{split}
\end{equation}
as shown in Figs. \ref{fig:Rpc}, \ref{fig:PnLISA}, \ref{fig:PnTaiji} and \ref{fig:PnTianQin}.

Thus, the sensitivity curves of the orthogonal modes $A$, $E$, $T$, constructed from the Relay combinations, are equal to those obtained from Michelson combinations, 
\begin{equation}
\label{eq:SnAET_UVW}
\begin{split}
S^{+|\times}_{n;A|E(U,V,W)}=& S^{+|\times}_{n;A|E(X,Y,Z)},\\
S^{+|\times}_{n;T(U,V,W)}=& S^{+|\times}_{n;T(X,Y,Z)},
\end{split}
\end{equation}
as shown in Fig. \ref{fig:SnAET}.

\subsection{Fully symmetric Sagnac combination ($\zeta$)}

The first-generation $\zeta$ combination is
\begin{equation}
\label{eq:zeta1}
\begin{split}
\zeta_{1} =& \mathcal{D}_{13}y_{21}+\mathcal{D}_{21}y_{32}+\mathcal{D}_{32}y_{13}\\
&-[\mathcal{D}_{12}y_{31}+\mathcal{D}_{23}y_{12}+\mathcal{D}_{31}y_{23}].
\end{split}
\end{equation}
The second-generation $\zeta$ combination is
\begin{equation}
\label{eq:zeta2}
\begin{split}
\zeta_{2} =& (\mathcal{D}_{1231321}-1)(\mathcal{D}_{1321}-1)\mathcal{D}_{23}X_{2}+ (1-\mathcal{D}_{12131})\Big[-\mathcal{D}_{321}\alpha_{2}+ \mathcal{D}_{31}\beta_{2}+ \mathcal{D}_{12} \gamma_{2}\Big]
\end{split}
\end{equation}

From Eqs. \eqref{eq:Ra} and \eqref{eq:zeta2},
the sky- and polarization-averaged response functions and noise PSDs of the $\zeta$ combination are
\begin{equation}
\label{eq:RZeta}
\begin{split}
&\frac{R^{+|\times}_{\zeta_{2}}}{192\sin^{2} \frac{3u}{2} \sin^{2} 2u\sin^{2}3u} =  \frac{R^{+|\times}_{\zeta_{1}}}{3}
= \frac{R^{+|\times}_{T(X_{0},Y_{0},Z_{0})}}{4\sin^{2}\frac{u}{2}}.
\end{split}
\end{equation}
and
\begin{equation}
\label{eq:PnZeta}
\begin{split}
&\frac{P_{n}^{\zeta_{2}}}{192\sin^{2} \frac{3u}{2} \sin^{2} 2u\sin^{2}3u} =  \frac{P_{n}^{\zeta_{1}}}{3}
= \frac{P_{n}^{T(X_{0},Y_{0},Z_{0})}}{4\sin^{2}\frac{u}{2}},
\end{split}
\end{equation}
as shown in Figs. \ref{fig:Rpc}, \ref{fig:PnLISA}, \ref{fig:PnTaiji} and \ref{fig:PnTianQin}.

The averaged response functions and noise PSDs of the $\zeta$ combination are proportional to those of the $T$ channel.
Consequently, its sensitivity curve is equal to that of the $T$ channel, 
\begin{equation}
\label{eq:SnZeta}
\begin{split}
S^{+|\times}_{n;\zeta}=& S^{+|\times}_{n;T(X,Y,Z)},
\end{split}
\end{equation}
as shown in Fig.~\ref{fig:Sn}.

\subsection{Comparison with unequal-arm sensitivities}

It is useful to compare the above equal-arm analytical results with recent numerical studies of unequal-arm TDI sensitivities.
A recent numerical investigation of several second-generation TDI configurations, using instantaneous armlengths taken from a realistic LISA-like orbit, found that the sensitivities of the science channels, i.e., the $A$ and $E$ channels, are nearly identical for different TDI configurations.
This result is consistent with our conclusion that the $A$ and $E$ sensitivities and the optimal sensitivity are independent of the specific TDI combination in the equilateral limit \cite{Wang:2025voa}.

The same study also showed that the $T$ channel can be more sensitive to the particular TDI configuration under unequal-arm conditions; in particular, the Michelson $T$ channel may no longer behave as a low-frequency null channel.
This difference is expected, because the present closed-form results rely on the instantaneous equilateral-constellation approximation, in which the cyclic symmetry among the three arms is preserved.
In the equal-arm and identical-noise limit considered here, the $A,E,T$ transformation diagonalizes the idealized noise covariance, and the nominal $T$ channel behaves as a low-frequency null-like channel.
Unequal armlengths break this symmetry and can induce residual correlations and mixing between the nominal science and null channels.

Therefore, our conclusion that the $T$-channel sensitivity is independent of the TDI configuration should be understood as an equal-arm result, while unequal-arm corrections can introduce configuration-dependent behavior.

\section{Conclusion}
\label{sec:conclusion}

Forthcoming space-based GW missions, such as LISA, TianQin, and Taiji, will employ TDI to suppress the overwhelming laser frequency noise and attain the required sensitivity.
The sensitivity curve of a TDI combination, defined as the ratio between its noise PSD and the sky‑ and polarization‑averaged response function, characterizes the overall performance of the detector.
Analytical expressions and accurate approximations for the averaged response function are useful for constructing sensitivity curves, comparing TDI observables, validating numerical calculations, and improving our understanding of the sensitivity behavior.

In this work, we introduced an inverse light‑path operator $\mathcal{P}_{i_{1}i_{2}i_{3}\ldots i_{n-1}i_{n}}$, which allows TDI combinations to be represented in a simple and compact form and provides a concise description of light propagation.
Under the instantaneous equal-arm, equilateral triangular-constellation approximation, we derived the sky‑ and polarization‑averaged response function for a general TDI combination,
and obtained the analytical averaged response functions for the second-generation TDI Michelson, Sagnac-type ($\alpha, \beta, \gamma$), Monitor, Beacon, Relay, and fully symmetric Sagnac ($\zeta$) combinations, as well as for the orthogonal $A, E, T$ channels formed from them.
We also presented the noise PSDs and the corresponding sensitivity curves for all of those combinations and their corresponding orthogonal $A, E, T$ channels.

Our analysis shows that:
(i) second‑generation TDIs have the same sensitivities as their first‑generation counterparts;
(ii) the sensitivity curves of the orthogonal $A, E, T$ channels and the optimal sensitivity are independent of the TDI generation and specific combination;
(iii) the averaged response functions, noise PSDs, and sensitivity curves of the $A$ and $E$ channels are equal,
while the $T$ channel exhibits a much lower response and sensitivity when $u=2\pi fL <3$ (that is, $f<0.057$ Hz for LISA, $f<0.048$ Hz for Taiji, and $f<0.83$ Hz for TianQin);
(iv) the averaged response function, noise PSD, and sensitivity curve of the fully symmetric Sagnac ($\zeta$) combination are proportional to those of the $T$ channel.

By matching the asymptotic behaviors in the low- and high-frequency limits of the analytical expressions, we further provided simple yet high‑accuracy approximate formulas that are convenient for practical use.
This asymptotic analysis shows that:
(v) sensitivity curves of the $A, E$ channels, as well as of all TDI combinations except the $(\alpha,\beta,\gamma)$ and $\zeta$ combinations, exhibit a flat section in the range $f_{n}<f\lesssim \frac{1.5}{2\pi L}$, where the noise-balance frequency $f_{n}$ marks the boundary between the proof-mass–dominated (low-frequency) and optical-path–dominated (high-frequency) noise regimes,
while the response-transition frequency $\sim \frac{1.5}{2\pi L}$ marks the boundary between the response function’s low- and high-frequency behaviors.

All conclusions listed above should be understood within the instantaneous equal-arm approximation.
In the SPA treatment, the common armlength $L$ may be evaluated at the stationary phase time, $L(t_f)$.
Thus, a slow common variation of the armlength can be incorporated at this level.
However, unequal-arm effects, i.e., corrections associated with different armlengths for different optical links, are not included in the closed-form expressions derived here.
The present results therefore provide analytical equal-arm benchmarks, while extensions to unequal-arm constellations are left for future work.

These analytical results and approximate formulas provide useful benchmarks for future studies on instrument optimization, sensitivity forecasts, data-analysis preparatory studies, and comparisons of TDI observables for space-based gravitational-wave detectors.

\acknowledgments
This work was supported by the National Natural Science Foundation of China under Grant No. 12605124.

\appendix
\renewcommand{\thefigure}{\Alph{section}.\arabic{figure}}
\setcounter{figure}{0}

\section{Numerical integration}
\label{app:num}

\begin{figure*}[htp]
	\centering
	\includegraphics[width=0.9\textwidth]{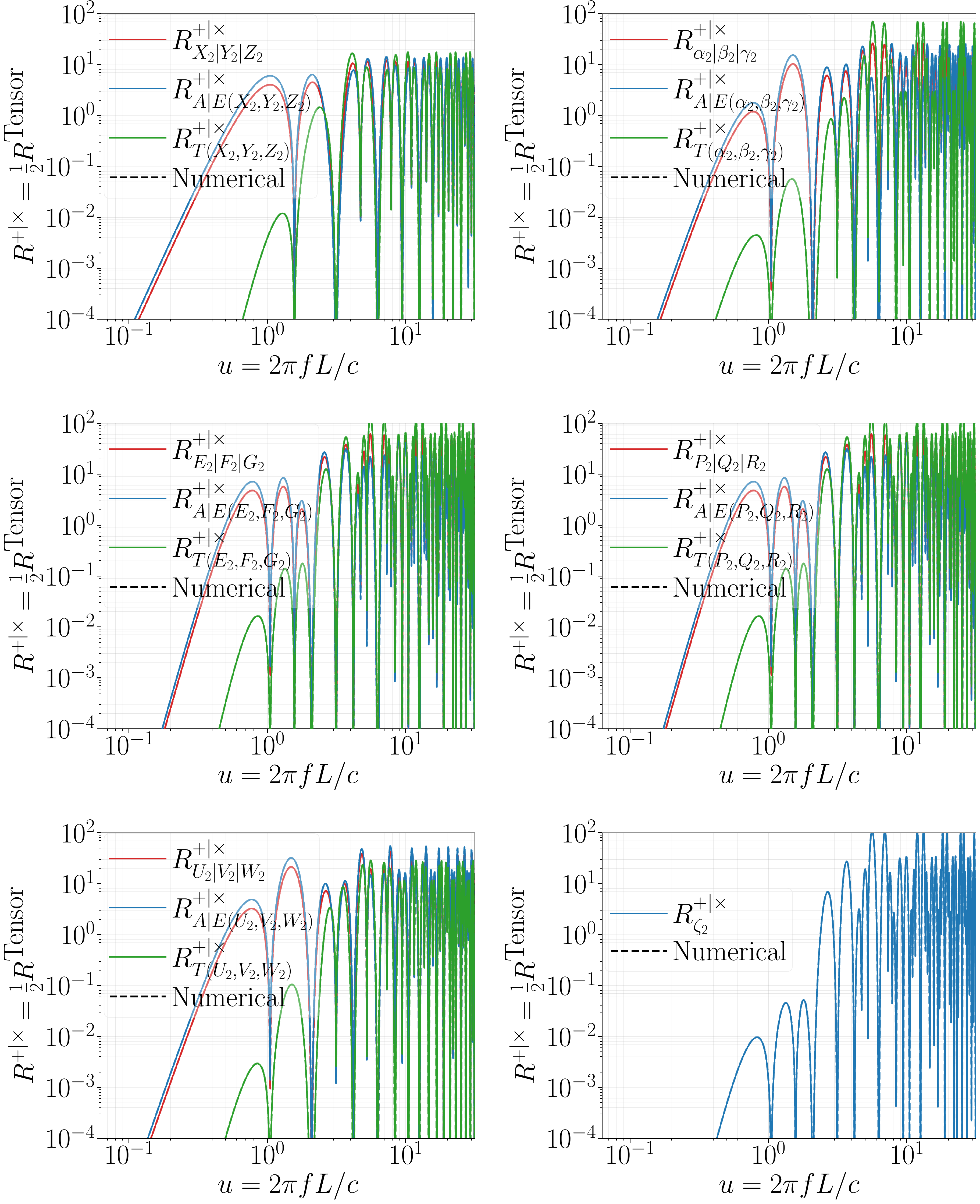}
	\caption{Comparison between the analytical expressions (solid lines) and numerical results (dashed lines) obtained via Monte Carlo integration using $10^{4}$ randomly sampled sky locations and polarization angles for Eqs. \eqref{eq:FA} and \eqref{eq:ra}.
	The two sets of results coincide, and the curves overlap entirely, confirming the validity of the analytical formulas.
	}
	\label{fig:RCompare}
\end{figure*}

To validate our analytical results, we directly apply Monte Carlo integration to Eqs. \eqref{eq:FA} and \eqref{eq:ra} to obtain numerical estimates. We find that when the number of randomly sampled sky locations and polarization angles exceeds $10^{4}$, the numerical values coincide with the analytical expressions within numerical precision. The comparison is shown in Fig. \ref{fig:RCompare}.

\section{Noise terms of $X_{2}$ and $\alpha_2$}
\label{app:noise}

The noise terms of the second-generation $X$ combination are
\begin{equation}
\label{eq:X2noise}
\begin{split}
X_{2}^{\text{noise}}
=&2 \big( -\mathcal{D}_{12} + \mathcal{D}_{1312} + \mathcal{D}_{131212} - \mathcal{D}_{12131312} \big) \boldsymbol{\hat{n}}_{12} \cdot \boldsymbol{\delta}_{21}\\
& + 2 \big( -\mathcal{D}_{13} + \mathcal{D}_{1213} + \mathcal{D}_{121313} - \mathcal{D}_{13121213} \big) \boldsymbol{\hat{n}}_{31} \cdot \boldsymbol{\delta}_{31} \\
& + \big( 1 + \mathcal{D}_{121} - \mathcal{D}_{131} - 2 \mathcal{D}_{13121} + \mathcal{D}_{1213131} - \mathcal{D}_{1312121} + \mathcal{D}_{121313121} \big) \boldsymbol{\hat{n}}_{12} \cdot \boldsymbol{\delta}_{12} \\
& + \big( 1 - \mathcal{D}_{121} + \mathcal{D}_{131} - 2 \mathcal{D}_{12131} - \mathcal{D}_{1213131} + \mathcal{D}_{1312121} + \mathcal{D}_{131212131} \big) \boldsymbol{\hat{n}}_{31} \cdot \boldsymbol{\delta}_{13} \\
& + \big( -1 + \mathcal{D}_{131} + \mathcal{D}_{13121} - \mathcal{D}_{1213131} \big) \mathcal{N}_{12} \\
& + \big( 1 - \mathcal{D}_{121} - \mathcal{D}_{12131} + \mathcal{D}_{1312121} \big) \mathcal{N}_{13} \\
& + \big( -\mathcal{D}_{12} + \mathcal{D}_{1312} + \mathcal{D}_{131212} - \mathcal{D}_{12131312} \big) \mathcal{N}_{21}  \\
&+ \big( \mathcal{D}_{13} - \mathcal{D}_{1213} - \mathcal{D}_{121313} + \mathcal{D}_{13121213} \big) \mathcal{N}_{31} \\
& + \big(\mathcal{D}_{131212131} -\mathcal{D}_{121313121}\big) p_{12}.
\end{split}
\end{equation}

The noise terms of the second-generation $\alpha$ combination are
\begin{equation}
\label{eq:alpha2noise}
\begin{split}
\alpha_{2}^{\text{noise}} 
=& \big( \mathcal{D}_{13} - \mathcal{D}_{123} - \mathcal{D}_{12313} + \mathcal{D}_{132123} - \mathcal{D}_{12313213} \\
&+ \mathcal{D}_{132123123} + \mathcal{D}_{13212312313} - \mathcal{D}_{123132132123} \big) \boldsymbol{\hat{n}}_{23} \cdot \boldsymbol{\delta}_{32} \\
& + \big( -\mathcal{D}_{13} + \mathcal{D}_{123} + \mathcal{D}_{12313} - \mathcal{D}_{132123} + \mathcal{D}_{12313213} \\
&- \mathcal{D}_{132123123} - \mathcal{D}_{13212312313} + \mathcal{D}_{123132132123} \big) \boldsymbol{\hat{n}}_{31} \cdot \boldsymbol{\delta}_{31} \\
& + \big( \mathcal{D}_{12} - \mathcal{D}_{132} - \mathcal{D}_{13212} + \mathcal{D}_{123132} - \mathcal{D}_{13212312} \\
&+ \mathcal{D}_{123132132} + \mathcal{D}_{12313213212} - \mathcal{D}_{132123123132} \big) \boldsymbol{\hat{n}}_{23} \cdot \boldsymbol{\delta}_{23} \\
& + \big( -\mathcal{D}_{12} + \mathcal{D}_{132} + \mathcal{D}_{13212} - \mathcal{D}_{123132} + \mathcal{D}_{13212312} \\
&- \mathcal{D}_{123132132} - \mathcal{D}_{12313213212} + \mathcal{D}_{132123123132} \big) \boldsymbol{\hat{n}}_{12} \cdot \boldsymbol{\delta}_{21} \\
& + \big( 1 - 2 \mathcal{D}_{1231} - \mathcal{D}_{1231321} + \mathcal{D}_{1321231} + 2 \mathcal{D}_{1321231231} - \mathcal{D}_{1231321321231} \big) \boldsymbol{\hat{n}}_{31} \cdot \boldsymbol{\delta}_{13} \\
& + \big( 1 - 2 \mathcal{D}_{1321} + \mathcal{D}_{1231321} - \mathcal{D}_{1321231} + 2 \mathcal{D}_{1231321321} - \mathcal{D}_{1321231231321} \big) \boldsymbol{\hat{n}}_{12} \cdot \boldsymbol{\delta}_{12} \\
& + \big( -1 + \mathcal{D}_{1321} + \mathcal{D}_{1321231} - \mathcal{D}_{1231321321} \big) \mathcal{N}_{12} \\
& + \big( 1 - \mathcal{D}_{1231} - \mathcal{D}_{1231321} + \mathcal{D}_{1321231231} \big) \mathcal{N}_{13} \\
& + \big( -\mathcal{D}_{12} + \mathcal{D}_{13212} + \mathcal{D}_{13212312} - \mathcal{D}_{12313213212} \big) \mathcal{N}_{23} \\
& + \big( \mathcal{D}_{13} - \mathcal{D}_{12313} - \mathcal{D}_{12313213} + \mathcal{D}_{13212312313} \big) \mathcal{N}_{32} \\
& + \big( -\mathcal{D}_{123} + \mathcal{D}_{132123} + \mathcal{D}_{132123123} - \mathcal{D}_{123132132123} \big) \mathcal{N}_{31} \\
& + \big( \mathcal{D}_{132} - \mathcal{D}_{123132} - \mathcal{D}_{123132132} + \mathcal{D}_{132123123132} \big) \mathcal{N}_{21} \\
& + \big( \mathcal{D}_{1321231231321} -\mathcal{D}_{1231321321231}\big) p_{12}.
\end{split}
\end{equation}

For space-based GW detectors, each spacecraft carries two optical benches, pointing toward the other two spacecraft.
Here $\boldsymbol{\delta}_{ij}$, $\mathcal{N}_{ij}$, and $p_{ij}$ denote the proof‑mass, optical‑path, and laser phase noise originating from the optical bench on $SC_{i}$ that points toward $SC_{j}$.
At linear order in the inter‑spacecraft velocities, $\mathcal{D}_{131212131} -\mathcal{D}_{121313121}=[\mathcal{D}_{13121},\mathcal{D}_{12131}]=0$ and $\mathcal{D}_{1321231231321} -\mathcal{D}_{1231321321231}=[\mathcal{D}_{1321231},\mathcal{D}_{1231321}]=0$; therefore, the laser phase‑noise terms are exactly canceled in the combinations $X_2$ and $\alpha_2$ owing to these commutators.

\section{Composite coefficients}
\label{app:coeff}

The composite coefficients for the second-generation Michelson combinations are
\begin{equation}
\label{}
\begin{split}
C_{1}^{X_{2}|Y_{2}|Z_{2}} =& 64 \sin^{2} u \sin^{2} 2u,\quad
C_{2}^{X_{2}|Y_{2}|Z_{2}} = 16 \sin u \sin^{3} 2u, \\
C_{3}^{X_{2}|Y_{2}|Z_{2}} =& -16 \sin u \sin^{3} 2u,\quad
C_{4}^{X_{2}|Y_{2}|Z_{2}} = 32 \sin^3 u \sin^{2} 2u,\\
C_{5}^{X_{2}|Y_{2}|Z_{2}} =& -32 \sin^{2} u \sin^{2} 2u.
\end{split}
\end{equation}
The composite coefficients for the corresponding orthogonal $A, E, T$ channels are
\begin{equation}
\label{}
\begin{split}
C_{1}^{A|E(X_{2},Y_{2},Z_{2})} =& 32(2+\cos u)\sin^{2} u \sin^{2} 2u,\quad
C_{2}^{A|E(X_{2},Y_{2},Z_{2})} = 16(1+2\cos u)\sin^{2} u \sin^{2} 2u,\\
C_{3}^{A|E(X_{2},Y_{2},Z_{2})} =& -16(2+\cos u)\sin^{2} u \sin^{2} 2u,\quad
C_{4}^{A|E(X_{2},Y_{2},Z_{2})} = 48\sin^{3} u \sin^{2} 2u,\\
C_{5}^{A|E(X_{2},Y_{2},Z_{2})} =& -16(1+2\cos u)\sin^{2} u \sin^{2} 2u,
\end{split}
\end{equation}
and
\begin{equation}
\label{}
\begin{split}
C_{1}^{T(X_{2},Y_{2},Z_{2})} =& 128\sin^{2}\frac{u}{2}\sin^{2}u\sin^{2}2u,\quad
C_{2}^{T(X_{2},Y_{2},Z_{2})} = -64\sin^{2}\frac{u}{2}\sin^{2}u\sin^{2}2u,\\
C_{3}^{T(X_{2},Y_{2},Z_{2})} =& 128\sin^{2}\frac{u}{2}\sin^{2}u\sin^{2}2u,\quad
C_{4}^{T(X_{2},Y_{2},Z_{2})} = 0,\\
C_{5}^{T(X_{2},Y_{2},Z_{2})} =& -128\sin^{2}\frac{u}{2}\sin^{2}u\sin^{2}2u.
\end{split}
\end{equation}

The composite coefficients for the second-generation ($\alpha, \beta, \gamma$) combinations are
\begin{equation}
\label{}
\begin{split}
C_{1}^{\alpha_{2}|\beta_{2}|\gamma_{2}} =& 96\sin^{2}\frac{3u}{2}\sin^{2}3u,\quad
C_{2}^{\alpha_{2}|\beta_{2}|\gamma_{2}} = -16(1+2\cos2u)\sin^{2}\frac{3u}{2}\sin^{2}3u,\\
C_{3}^{\alpha_{2}|\beta_{2}|\gamma_{2}} =& 32(2\cos u+\cos2u)\sin^{2}\frac{3u}{2}\sin^{2}3u,\quad
C_{4}^{\alpha_{2}|\beta_{2}|\gamma_{2}} = 128\sin^{2}\frac{u}{2}\sin u\sin^{2}\frac{3u}{2}\sin^{2}3u,\\
C_{5}^{\alpha_{2}|\beta_{2}|\gamma_{2}} =& -32(1+2\cos u)\sin^{2}\frac{3u}{2}\sin^{2}3u.
\end{split}
\end{equation}
The composite coefficients for the corresponding orthogonal $A, E, T$ channels are
\begin{equation}
\label{}
\begin{split}
C_{1}^{A|E(\alpha_{2},\beta_{2},\gamma_{2})} =& 128(2+\cos u)\sin^{2}\frac{u}{2}\sin^{2}\frac{3u}{2}\sin^{2}3u,\\
C_{2}^{A|E(\alpha_{2},\beta_{2},\gamma_{2})} =& 64(1+2\cos u)\sin^{2}\frac{u}{2}\sin^{2}\frac{3u}{2}\sin^{2}3u,\\
C_{3}^{A|E(\alpha_{2},\beta_{2},\gamma_{2})} =& -64(2+\cos u)\sin^{2}\frac{u}{2}\sin^{2}\frac{3u}{2}\sin^{2}3u,\\
C_{4}^{A|E(\alpha_{2},\beta_{2},\gamma_{2})} =& 192\sin^{2}\frac{u}{2}\sin u\sin^{2}\frac{3u}{2}\sin^{2}3u,\\
C_{5}^{A|E(\alpha_{2},\beta_{2},\gamma_{2})} =& -64(1+2\cos u)\sin^{2}\frac{u}{2}\sin^{2}\frac{3u}{2}\sin^{2}3u,
\end{split}
\end{equation}
and
\begin{equation}
\label{}
\begin{split}
C_{1}^{T(\alpha_{2},\beta_{2},\gamma_{2})} =& 32(1+2\cos u)^{2}\sin^{2}\frac{3u}{2}\sin^{2}3u,\\
C_{2}^{T(\alpha_{2},\beta_{2},\gamma_{2})} =& -16(1+2\cos u)^{2}\sin^{2}\frac{3u}{2}\sin^{2}3u,\\
C_{3}^{T(\alpha_{2},\beta_{2},\gamma_{2})} =& 32(1+2\cos u)^{2}\sin^{2}\frac{3u}{2}\sin^{2}3u,\\
C_{4}^{T(\alpha_{2},\beta_{2},\gamma_{2})} =& 0,\\
C_{5}^{T(\alpha_{2},\beta_{2},\gamma_{2})} =& -32(1+2\cos u)^{2}\sin^{2}\frac{3u}{2}\sin^{2}3u.
\end{split}
\end{equation}

The composite coefficients for the second-generation Monitor combinations are
\begin{equation}
\label{}
\begin{split}
C_{1}^{E_{2}|F_{2}|G_{2}} =& 512(3+2\cos u)\sin^{2}\frac{u}{2}\sin^{2}\frac{3u}{2}\sin^{2}2u\sin^{2}3u,\\
C_{2}^{E_{2}|F_{2}|G_{2}} =& -256\sin^{2}\frac{u}{2}\sin^{2}\frac{3u}{2}\sin^{2}2u\sin^{2}3u,\\
C_{3}^{E_{2}|F_{2}|G_{2}} =& 256\sin^{2}u\sin^{2}\frac{3u}{2}\sin^{2}2u\sin^{2}3u,\\
C_{4}^{E_{2}|F_{2}|G_{2}} =& 512\sin^{2}\frac{u}{2}\sin u\sin^{2}\frac{3u}{2}\sin^{2}2u\sin^{2}3u,\\
C_{5}^{E_{2}|F_{2}|G_{2}} =& -512\sin^{2}u\sin^{2}\frac{3u}{2}\sin^{2}2u\sin^{2}3u.
\end{split}
\end{equation}
The composite coefficients for the corresponding orthogonal $A, E, T$ channels are
\begin{equation}
\label{}
\begin{split}
C_{1}^{A|E(E_{2},F_{2},G_{2})} =& 512(2+\cos u)\sin^{2}\frac{u}{2}\sin^{2}\frac{3u}{2}\sin^{2}2u\sin^{2}3u,\\
C_{2}^{A|E(E_{2},F_{2},G_{2})} =& 256(1+2\cos u)\sin^{2}\frac{u}{2}\sin^{2}\frac{3u}{2}\sin^{2}2u\sin^{2}3u,\\
C_{3}^{A|E(E_{2},F_{2},G_{2})} =& -256(2+\cos u)\sin^{2}\frac{u}{2}\sin^{2}\frac{3u}{2}\sin^{2}2u\sin^{2}3u,\\
C_{4}^{A|E(E_{2},F_{2},G_{2})} =& 768\sin^{2}\frac{u}{2}\sin u\sin^{2}\frac{3u}{2}\sin^{2}2u\sin^{2}3u,\\
C_{5}^{A|E(E_{2},F_{2},G_{2})} =& -256(1+2\cos u)\sin^{2}\frac{u}{2}\sin^{2}\frac{3u}{2}\sin^{2}2u\sin^{2}3u,
\end{split}
\end{equation}
and
\begin{equation}
\label{}
\begin{split}
	C_{1}^{T(E_{2},F_{2},G_{2})} =& 512(5+4\cos u)\sin^{2}\frac{u}{2}\sin^{2}\frac{3u}{2}\sin^{2}2u\sin^{2}3u,\\
	C_{2}^{T(E_{2},F_{2},G_{2})} =& -256(5+4\cos u)\sin^{2}\frac{u}{2}\sin^{2}\frac{3u}{2}\sin^{2}2u\sin^{2}3u,\\
	C_{3}^{T(E_{2},F_{2},G_{2})} =& 512(5+4\cos u)\sin^{2}\frac{u}{2}\sin^{2}\frac{3u}{2}\sin^{2}2u\sin^{2}3u,\\
	C_{4}^{T(E_{2},F_{2},G_{2})} =& 0,\\
	C_{5}^{T(E_{2},F_{2},G_{2})} =& -512(5+4\cos u)\sin^{2}\frac{u}{2}\sin^{2}\frac{3u}{2}\sin^{2}2u\sin^{2}3u.
\end{split}
\end{equation}

The composite coefficients for the second-generation Beacon combinations are
equal to those for the second-generation Monitor combinations.

The composite coefficients for the second-generation Relay combinations are

\begin{equation}
\label{}
\begin{split}
	C_{1}^{U_{2}|V_{2}|W_{2}} =& 64 \sin^{2} u \sin^{2} 2u,~
	C_{2}^{U_{2}|V_{2}|W_{2}} = 16 \sin u \sin^{3} 2u,~ 
	C_{3}^{U_{2}|V_{2}|W_{2}} = -16 \sin u \sin^{3} 2u, \\
	C_{4}^{U_{2}|V_{2}|W_{2}} =& 32 \sin^3 u \sin^{2} 2u,~
	C_{5}^{U_{2}|V_{2}|W_{2}} = -32 \sin^{2} u \sin^{2} 2u,
\end{split}
\end{equation}
The composite coefficients for the corresponding orthogonal $A, E, T$ channels are
\begin{equation}
\begin{split}
	C_{1}^{A|E(U_{2},V_{2},W_{2})} =& 32(2+\cos u)\sin^{2} u \sin^{2} 2u,~
	C_{2}^{A|E(U_{2},V_{2},W_{2})} = 16(1+2\cos u)\sin^{2} u \sin^{2} 2u,\\
	C_{3}^{A|E(U_{2},V_{2},W_{2})} =& -16(2+\cos u)\sin^{2} u \sin^{2} 2u,~
	C_{4}^{A|E(U_{2},V_{2},W_{2})} = 48\sin^{3} u \sin^{2} 2u,\\
	C_{5}^{A|E(U_{2},V_{2},W_{2})} =&-16(1+2\cos u)\sin^{2} u \sin^{2} 2u,
\end{split}
\end{equation}
and
\begin{equation}
\begin{split}
	C_{1}^{T(U_{2},V_{2},W_{2})} =& 128\sin^{2}\frac{u}{2}\sin^{2}u\sin^{2}2u,~
	C_{2}^{T(U_{2},V_{2},W_{2})} = -64\sin^{2}\frac{u}{2}\sin^{2}u\sin^{2}2u,\\
	C_{3}^{T(U_{2},V_{2},W_{2})} =& 128\sin^{2}\frac{u}{2}\sin^{2}u\sin^{2}2u,~
	C_{4}^{T(U_{2},V_{2},W_{2})} = 0,\\
	C_{5}^{T(U_{2},V_{2},W_{2})} =& -128\sin^{2}\frac{u}{2}\sin^{2}u\sin^{2}2u.
\end{split}
\end{equation}

The composite coefficients for the second-generation $\zeta$ combination are

\begin{equation}
\label{}
\begin{split}
	C_{1}^{\zeta_{2}} =& 384 \sin^{2}\frac{3u}{2}\sin^{2}2u\sin^{2}3u,~
	C_{2}^{\zeta_{2}} = -192 \sin^{2}\frac{3u}{2}\sin^{2}2u\sin^{2}3u,\\
	C_{3}^{\zeta_{2}} =& 384 \sin^{2}\frac{3u}{2}\sin^{2}2u\sin^{2}3u,~
	C_{4}^{\zeta_{2}} = 0,~
	C_{5}^{\zeta_{2}} = -384 \sin^{2}\frac{3u}{2}\sin^{2}2u\sin^{2}3u.
\end{split}
\end{equation}

\bibliographystyle{JHEP}
\bibliography{ref}

@article{LIGOScientific:2016aoc,
    author = "Abbott, B. P. and others",
    collaboration = "LIGO Scientific, Virgo",
    title = "{Observation of Gravitational Waves from a Binary Black Hole Merger}",
    eprint = "1602.03837",
    archivePrefix = "arXiv",
    primaryClass = "gr-qc",
    reportNumber = "LIGO-P150914",
    doi = "10.1103/PhysRevLett.116.061102",
    journal = "Phys. Rev. Lett.",
    volume = "116",
    number = "6",
    pages = "061102",
    year = "2016"
}

@article{LIGOScientific:2016emj,
    author = "Abbott, B. P. and others",
    collaboration = "LIGO Scientific, Virgo",
    title = "{GW150914: The Advanced LIGO Detectors in the Era of First Discoveries}",
    eprint = "1602.03838",
    archivePrefix = "arXiv",
    primaryClass = "gr-qc",
    reportNumber = "LIGO-P1500237",
    doi = "10.1103/PhysRevLett.116.131103",
    journal = "Phys. Rev. Lett.",
    volume = "116",
    number = "13",
    pages = "131103",
    year = "2016"
}

@article{LIGOScientific:2016vlm,
    author = "Abbott, B. P. and others",
    collaboration = "LIGO Scientific, Virgo",
    title = "{Properties of the Binary Black Hole Merger GW150914}",
    eprint = "1602.03840",
    archivePrefix = "arXiv",
    primaryClass = "gr-qc",
    reportNumber = "LIGO-P1500218",
    doi = "10.1103/PhysRevLett.116.241102",
    journal = "Phys. Rev. Lett.",
    volume = "116",
    number = "24",
    pages = "241102",
    year = "2016"
}

@article{LIGOScientific:2018mvr,
    author = "Abbott, B. P. and others",
    collaboration = "LIGO Scientific, Virgo",
    title = "{GWTC-1: A Gravitational-Wave Transient Catalog of Compact Binary Mergers Observed by LIGO and Virgo during the First and Second Observing Runs}",
    eprint = "1811.12907",
    archivePrefix = "arXiv",
    primaryClass = "astro-ph.HE",
    reportNumber = "LIGO-P1800307",
    doi = "10.1103/PhysRevX.9.031040",
    journal = "Phys. Rev. X",
    volume = "9",
    number = "3",
    pages = "031040",
    year = "2019"
}

@article{LIGOScientific:2020ibl,
    author = "Abbott, R. and others",
    collaboration = "LIGO Scientific, Virgo",
    title = "{GWTC-2: Compact Binary Coalescences Observed by LIGO and Virgo During the First Half of the Third Observing Run}",
    eprint = "2010.14527",
    archivePrefix = "arXiv",
    primaryClass = "gr-qc",
    reportNumber = "P2000061",
    doi = "10.1103/PhysRevX.11.021053",
    journal = "Phys. Rev. X",
    volume = "11",
    pages = "021053",
    year = "2021"
}

@article{KAGRA:2021vkt,
    author = "Abbott, R. and others",
    collaboration = "KAGRA, VIRGO, LIGO Scientific",
    title = "{GWTC-3: Compact Binary Coalescences Observed by LIGO and Virgo during the Second Part of the Third Observing Run}",
    eprint = "2111.03606",
    archivePrefix = "arXiv",
    primaryClass = "gr-qc",
    reportNumber = "LIGO-P2000318",
    doi = "10.1103/PhysRevX.13.041039",
    journal = "Phys. Rev. X",
    volume = "13",
    number = "4",
    pages = "041039",
    year = "2023"
}

@article{LIGOScientific:2017ync,
    author = "Abbott, B. P. and others",
    title = "{Multi-messenger Observations of a Binary Neutron Star Merger}",
    eprint = "1710.05833",
    archivePrefix = "arXiv",
    primaryClass = "astro-ph.HE",
    reportNumber = "LIGO-P1700294, VIR-0802A-17, FERMILAB-PUB-17-478-A-AE-CD",
    doi = "10.3847/2041-8213/aa91c9",
    journal = "Astrophys. J. Lett.",
    volume = "848",
    number = "2",
    pages = "L12",
    year = "2017"
}

@article{LIGOScientific:2025slb,
    author = "Abac, A. G. and others",
    collaboration = "LIGO Scientific, VIRGO, KAGRA",
    title = "{GWTC-4.0: Updating the Gravitational-Wave Transient Catalog with Observations from the First Part of the Fourth LIGO-Virgo-KAGRA Observing Run}",
    eprint = "2508.18082",
    archivePrefix = "arXiv",
    primaryClass = "gr-qc",
    reportNumber = "LIGO-P2400386",
    doi = "10.3847/2041-8213/ae2c74",
    journal = "Astrophys. J. Lett.",
    volume = "1004",
    number = "2",
    pages = "L22",
    year = "2026"
}

@article{Harry:2010zz,
    author = "Harry, Gregory M.",
    editor = "Marka, Zsuzsa and Marka, Szabolcs",
    collaboration = "LIGO Scientific",
    title = "{Advanced LIGO: The next generation of gravitational wave detectors}",
    doi = "10.1088/0264-9381/27/8/084006",
    journal = "Class. Quant. Grav.",
    volume = "27",
    pages = "084006",
    year = "2010"
}

@article{TheLIGOScientific:2014jea,
    author = "Aasi, J. and others",
    collaboration = "LIGO Scientific",
    title = "{Advanced LIGO}",
    eprint = "1411.4547",
    archivePrefix = "arXiv",
    primaryClass = "gr-qc",
    doi = "10.1088/0264-9381/32/7/074001",
    journal = "Class. Quant. Grav.",
    volume = "32",
    pages = "074001",
    year = "2015"
}

@article{TheVirgo:2014hva,
    author = "Acernese, F. and others",
    collaboration = "VIRGO",
    title = "{Advanced Virgo: a second-generation interferometric gravitational wave detector}",
    eprint = "1408.3978",
    archivePrefix = "arXiv",
    primaryClass = "gr-qc",
    doi = "10.1088/0264-9381/32/2/024001",
    journal = "Class. Quant. Grav.",
    volume = "32",
    number = "2",
    pages = "024001",
    year = "2015"
}

@article{Somiya:2011np,
    author = "Somiya, Kentaro",
    editor = "Hannam, Mark and Sutton, Patrick and Hild, Stefan and van den Broeck, Chris",
    collaboration = "KAGRA",
    title = "{Detector configuration of KAGRA: The Japanese cryogenic gravitational-wave detector}",
    eprint = "1111.7185",
    archivePrefix = "arXiv",
    primaryClass = "gr-qc",
    doi = "10.1088/0264-9381/29/12/124007",
    journal = "Class. Quant. Grav.",
    volume = "29",
    pages = "124007",
    year = "2012"
}

@article{Aso:2013eba,
    author = "Aso, Yoichi and Michimura, Yuta and Somiya, Kentaro and Ando, Masaki and Miyakawa, Osamu and Sekiguchi, Takanori and Tatsumi, Daisuke and Yamamoto, Hiroaki",
    collaboration = "KAGRA",
    title = "{Interferometer design of the KAGRA gravitational wave detector}",
    eprint = "1306.6747",
    archivePrefix = "arXiv",
    primaryClass = "gr-qc",
    doi = "10.1103/PhysRevD.88.043007",
    journal = "Phys. Rev. D",
    volume = "88",
    number = "4",
    pages = "043007",
    year = "2013"
}

@article{Danzmann:1997hm,
    author = "Danzmann, K.",
    title = "{LISA: An ESA cornerstone mission for a gravitational wave observatory}",
    doi = "10.1088/0264-9381/14/6/002",
    journal = "Class. Quant. Grav.",
    volume = "14",
    pages = "1399--1404",
    year = "1997"
}

@article{LISA:2017pwj,
    author = "Amaro-Seoane, Pau and others",
    collaboration = "LISA",
    title = "{Laser Interferometer Space Antenna}",
    eprint = "1702.00786",
    archivePrefix = "arXiv",
    primaryClass = "astro-ph.IM",
    month = "2",
    year = "2017"
}

@article{Luo:2015ght,
    author = "Luo, Jun and others",
    collaboration = "TianQin",
    title = "{TianQin: a space-borne gravitational wave detector}",
    eprint = "1512.02076",
    archivePrefix = "arXiv",
    primaryClass = "astro-ph.IM",
    doi = "10.1088/0264-9381/33/3/035010",
    journal = "Class. Quant. Grav.",
    volume = "33",
    number = "3",
    pages = "035010",
    year = "2016"
}

@article{Hu:2017mde,
    author = "Hu, Wen-Rui and Wu, Yue-Liang",
    title = "{The Taiji Program in Space for gravitational wave physics and the nature of gravity}",
    doi = "10.1093/nsr/nwx116",
    journal = "Natl. Sci. Rev.",
    volume = "4",
    number = "5",
    pages = "685--686",
    year = "2017"
}

@article{Kawamura:2011zz,
    author = "Kawamura, Seiji and others",
    editor = "Buchman, Sasha and Sun, Ke-Xun",
    title = "{The Japanese space gravitational wave antenna: DECIGO}",
    doi = "10.1088/0264-9381/28/9/094011",
    journal = "Class. Quant. Grav.",
    volume = "28",
    pages = "094011",
    year = "2011"
}

@article{Abbott:2016nmj,
    author = "Abbott, B. P. and others",
    collaboration = "LIGO Scientific, Virgo",
    title = "{GW151226: Observation of Gravitational Waves from a 22-Solar-Mass Binary Black Hole Coalescence}",
    eprint = "1606.04855",
    archivePrefix = "arXiv",
    primaryClass = "gr-qc",
    reportNumber = "LIGO-P151226",
    doi = "10.1103/PhysRevLett.116.241103",
    journal = "Phys. Rev. Lett.",
    volume = "116",
    number = "24",
    pages = "241103",
    year = "2016"
}

@article{Abbott:2017vtc,
    author = "Abbott, Benjamin P. and others",
    collaboration = "LIGO Scientific, VIRGO",
    title = "{GW170104: Observation of a 50-Solar-Mass Binary Black Hole Coalescence at Redshift 0.2}",
    eprint = "1706.01812",
    archivePrefix = "arXiv",
    primaryClass = "gr-qc",
    reportNumber = "LIGO-P170104",
    doi = "10.1103/PhysRevLett.118.221101",
    journal = "Phys. Rev. Lett.",
    volume = "118",
    number = "22",
    pages = "221101",
    year = "2017",
    note = "[Erratum: Phys.Rev.Lett. 121, 129901 (2018)]"
}

@article{Abbott:2017oio,
    author = "Abbott, B. P. and others",
    collaboration = "LIGO Scientific, Virgo",
    title = "{GW170814: A Three-Detector Observation of Gravitational Waves from a Binary Black Hole Coalescence}",
    eprint = "1709.09660",
    archivePrefix = "arXiv",
    primaryClass = "gr-qc",
    doi = "10.1103/PhysRevLett.119.141101",
    journal = "Phys. Rev. Lett.",
    volume = "119",
    number = "14",
    pages = "141101",
    year = "2017"
}

@article{TheLIGOScientific:2017qsa,
    author = "Abbott, B. P. and others",
    collaboration = "LIGO Scientific, Virgo",
    title = "{GW170817: Observation of Gravitational Waves from a Binary Neutron Star Inspiral}",
    eprint = "1710.05832",
    archivePrefix = "arXiv",
    primaryClass = "gr-qc",
    reportNumber = "LIGO-P170817",
    doi = "10.1103/PhysRevLett.119.161101",
    journal = "Phys. Rev. Lett.",
    volume = "119",
    number = "16",
    pages = "161101",
    year = "2017"
}

@article{Abbott:2017gyy,
    author = "Abbott, B. . P. . and others",
    collaboration = "LIGO Scientific, Virgo",
    title = "{GW170608: Observation of a 19-solar-mass Binary Black Hole Coalescence}",
    eprint = "1711.05578",
    archivePrefix = "arXiv",
    primaryClass = "astro-ph.HE",
    reportNumber = "LIGO-DOCUMENT-P170608-V8",
    doi = "10.3847/2041-8213/aa9f0c",
    journal = "Astrophys. J. Lett.",
    volume = "851",
    pages = "L35",
    year = "2017"
}

@article{Tinto:1999yr,
    author = "Tinto, Massimo and Armstrong, J. W.",
    title = "{Cancellation of laser noise in an unequal-arm interferometer detector of gravitational radiation}",
    doi = "10.1103/PhysRevD.59.102003",
    journal = "Phys. Rev. D",
    volume = "59",
    pages = "102003",
    year = "1999"
}

@article{Armstrong:1999,
doi = {10.1086/308110},
url = {https://doi.org/10.1086/308110},
year = {1999},
month = {dec},
publisher = {},
volume = {527},
number = {2},
pages = {814},
author = {Armstrong, J. W. and Estabrook, F. B. and Tinto, Massimo},
title = {Time-Delay Interferometry for Space-based Gravitational Wave Searches},
journal = {The Astrophysical Journal},
}

@article{Estabrook:2000ef,
    author = "Estabrook, F. B. and Tinto, Massimo and Armstrong, J. W.",
    title = "{Time delay analysis of LISA gravitational wave data: Elimination of spacecraft motion effects}",
    doi = "10.1103/PhysRevD.62.042002",
    journal = "Phys. Rev. D",
    volume = "62",
    pages = "042002",
    year = "2000"
}

@article{Dhurandhar:2001tct,
    author = "Dhurandhar, S. V. and Rajesh Nayak, K. and Vinet, J. Y.",
    title = "{Algebraic approach to time-delay data analysis for LISA}",
    eprint = "gr-qc/0112059",
    archivePrefix = "arXiv",
    doi = "10.1103/PhysRevD.65.102002",
    journal = "Phys. Rev. D",
    volume = "65",
    number = "10",
    pages = "102002",
    year = "2002"
}

@article{Tinto:2020fcc,
    author = "Tinto, Massimo and Dhurandhar, Sanjeev V.",
    title = "{Time-delay interferometry}",
    doi = "10.1007/s41114-020-00029-6",
    journal = "Living Rev. Rel.",
    volume = "24",
    number = "1",
    pages = "1",
    year = "2021"
}

@article{Tinto:2003vj,
    author = "Tinto, Massimo and Estabrook, Frank B. and Armstrong, J. W.",
    title = "{Time delay interferometry with moving spacecraft arrays}",
    eprint = "gr-qc/0310017",
    archivePrefix = "arXiv",
    doi = "10.1103/PhysRevD.69.082001",
    journal = "Phys. Rev. D",
    volume = "69",
    pages = "082001",
    year = "2004"
}

@article{Nayak:2005un,
    author = "Nayak, K. R. and Vinet, J. Y.",
    editor = "Jennrich, O.",
    title = "{Algebraic approach to time-delay data analysis: Orbiting case}",
    doi = "10.1088/0264-9381/22/10/040",
    journal = "Class. Quant. Grav.",
    volume = "22",
    pages = "S437--S443",
    year = "2005"
}

@article{Vallisneri:2005ji,
    author = "Vallisneri, Michele",
    title = "{Geometric time delay interferometry}",
    eprint = "gr-qc/0504145",
    archivePrefix = "arXiv",
    doi = "10.1103/PhysRevD.76.109903",
    journal = "Phys. Rev. D",
    volume = "72",
    pages = "042003",
    year = "2005",
    note = "[Erratum: Phys.Rev.D 76, 109903 (2007)]"
}

@article{Muratore:2020mdf,
    author = "Muratore, Martina and Vetrugno, Daniele and Vitale, Stefano",
    title = "{Revisitation of time delay interferometry combinations that suppress laser noise in LISA}",
    eprint = "2001.11221",
    archivePrefix = "arXiv",
    primaryClass = "astro-ph.IM",
    doi = "10.1088/1361-6382/ab9d5b",
    journal = "Class. Quant. Grav.",
    volume = "37",
    number = "18",
    pages = "185019",
    year = "2020"
}

@article{Muratore:2021uqj,
    author = "Muratore, Martina and Vetrugno, Daniele and Vitale, Stefano and Hartwig, Olaf",
    title = "{Time delay interferometry combinations as instrument noise monitors for LISA}",
    eprint = "2108.02738",
    archivePrefix = "arXiv",
    primaryClass = "gr-qc",
    doi = "10.1103/PhysRevD.105.023009",
    journal = "Phys. Rev. D",
    volume = "105",
    number = "2",
    pages = "023009",
    year = "2022"
}

@article{Tinto:2022zmf,
    author = "Tinto, Massimo and Dhurandhar, Sanjeev and Malakar, Dishari",
    title = "{Second-generation time-delay interferometry}",
    eprint = "2212.05967",
    archivePrefix = "arXiv",
    primaryClass = "gr-qc",
    doi = "10.1103/PhysRevD.107.082001",
    journal = "Phys. Rev. D",
    volume = "107",
    number = "8",
    pages = "082001",
    year = "2023"
}

@article{Larson:1999we,
    author = "Larson, Shane L. and Hiscock, William A. and Hellings, Ronald W.",
    title = "{Sensitivity curves for spaceborne gravitational wave interferometers}",
    eprint = "gr-qc/9909080",
    archivePrefix = "arXiv",
    reportNumber = "MSUPHY-99-04",
    doi = "10.1103/PhysRevD.62.062001",
    journal = "Phys. Rev. D",
    volume = "62",
    pages = "062001",
    year = "2000"
}

@article{Robson:2018ifk,
    author = "Robson, Travis and Cornish, Neil J. and Liu, Chang",
    title = "{The construction and use of LISA sensitivity curves}",
    eprint = "1803.01944",
    archivePrefix = "arXiv",
    primaryClass = "astro-ph.HE",
    doi = "10.1088/1361-6382/ab1101",
    journal = "Class. Quant. Grav.",
    volume = "36",
    number = "10",
    pages = "105011",
    year = "2019"
}

@article{Zhang:2021kkh,
    author = "Zhang, Chunyu and Gong, Yungui and Zhang, Chao",
    title = "{Parameter estimation for space-based gravitational wave detectors with ringdown signals}",
    eprint = "2105.11279",
    archivePrefix = "arXiv",
    primaryClass = "gr-qc",
    doi = "10.1103/PhysRevD.104.083038",
    journal = "Phys. Rev. D",
    volume = "104",
    number = "8",
    pages = "083038",
    year = "2021"
}

@article{Zhang:2021wwd,
    author = "Zhang, Chunyu and Gong, Yungui and Zhang, Chao",
    title = "{Source localizations with the network of space-based gravitational wave detectors}",
    eprint = "2112.02299",
    archivePrefix = "arXiv",
    primaryClass = "gr-qc",
    doi = "10.1103/PhysRevD.106.024004",
    journal = "Phys. Rev. D",
    volume = "106",
    number = "2",
    pages = "024004",
    year = "2022"
}

@article{Zhang:2020khm,
    author = "Zhang, Chunyu and Gao, Qing and Gong, Yungui and Wang, Bin and Weinstein, Alan J. and Zhang, Chao",
    title = "{Full analytical formulas for frequency response of space-based gravitational wave detectors}",
    eprint = "2003.01441",
    archivePrefix = "arXiv",
    primaryClass = "gr-qc",
    doi = "10.1103/PhysRevD.101.124027",
    journal = "Phys. Rev. D",
    volume = "101",
    number = "12",
    pages = "124027",
    year = "2020"
}

@article{Zhang:2019oet,
    author = "Zhang, Chunyu and Gao, Qing and Gong, Yungui and Liang, Dicong and Weinstein, Alan J. and Zhang, Chao",
    title = "{Frequency response of time-delay interferometry for space-based gravitational wave antenna}",
    eprint = "1906.10901",
    archivePrefix = "arXiv",
    primaryClass = "gr-qc",
    doi = "10.1103/PhysRevD.100.064033",
    journal = "Phys. Rev. D",
    volume = "100",
    number = "6",
    pages = "064033",
    year = "2019"
}

@article{Wang:2021jsv,
    author = "Wang, Pan-Pan and Tan, Yu-Jie and Qian, Wei-Liang and Shao, Cheng-Gang",
    title = "{Sensitivity functions of spaceborne gravitational wave detectors for arbitrary time-delay interferometry combinations}",
    doi = "10.1103/PhysRevD.103.063021",
    journal = "Phys. Rev. D",
    volume = "103",
    number = "6",
    pages = "063021",
    year = "2021"
}

@article{Cornish:2002rt,
    author = "Cornish, Neil J. and Rubbo, Louis J.",
    title = "{The LISA response function}",
    eprint = "gr-qc/0209011",
    archivePrefix = "arXiv",
    doi = "10.1103/PhysRevD.67.029905",
    journal = "Phys. Rev. D",
    volume = "67",
    pages = "022001",
    year = "2003",
    note = "[Erratum: Phys.Rev.D 67, 029905 (2003)]"
}

@article{Prince:2002hp,
    author = "Prince, Thomas A. and Tinto, Massimo and Larson, Shane L. and Armstrong, J. W.",
    title = "{The LISA optimal sensitivity}",
    eprint = "gr-qc/0209039",
    archivePrefix = "arXiv",
    doi = "10.1103/PhysRevD.66.122002",
    journal = "Phys. Rev. D",
    volume = "66",
    pages = "122002",
    year = "2002"
}

@article{Ruan:2020smc,
    author = "Ruan, Wen-Hong and Liu, Chang and Guo, Zong-Kuan and Wu, Yue-Liang and Cai, Rong-Gen",
    title = "{The LISA-Taiji network}",
    eprint = "2002.03603",
    archivePrefix = "arXiv",
    primaryClass = "gr-qc",
    doi = "10.1038/s41550-019-1008-4",
    journal = "Nature Astron.",
    volume = "4",
    pages = "108--109",
    year = "2020"
}

@article{Wang:2025voa,
    author = "Wang, Gang",
    title = "{Correlation and data-analysis distinctiveness of time-delay interferometry configurations}",
    eprint = "2507.18397",
    archivePrefix = "arXiv",
    primaryClass = "gr-qc",
    doi = "10.1103/1tvy-dztc",
    journal = "Phys. Rev. D",
    volume = "113",
    number = "12",
    pages = "124072",
    year = "2026"
}

@article{Marsat:2020rtl,
    author = "Marsat, Sylvain and Baker, John G. and Dal Canton, Tito",
    title = "{Exploring the Bayesian parameter estimation of binary black holes with LISA}",
    eprint = "2003.00357",
    archivePrefix = "arXiv",
    primaryClass = "gr-qc",
    doi = "10.1103/PhysRevD.103.083011",
    journal = "Phys. Rev. D",
    volume = "103",
    number = "8",
    pages = "083011",
    year = "2021"
}

\end{document}